\documentclass[journal]{IEEEtran}

\usepackage[utf8]{inputenc}
\usepackage[T1]{fontenc}
\usepackage[english]{babel}

\usepackage{mathptmx}  % Times font for text + math

\usepackage{graphicx}
\usepackage{caption}
\usepackage{pbalance}
\usepackage{multirow}
\usepackage{tabularx}
\usepackage{array}
\usepackage{wrapfig}
\usepackage{float}
\usepackage{adjustbox}
\usepackage{colortbl}
\usepackage[table,xcdraw]{xcolor}
\usepackage{xurl}
\newcolumntype{Y}{>{\raggedright\arraybackslash}X} 
\usepackage{amssymb} 
\usepackage{makecell}
\usepackage{wasysym}
\usepackage{tcolorbox}
\usepackage{booktabs}
\usepackage{dblfloatfix}
\definecolor{lightgreen}{RGB}{200,255,200}
\definecolor{verylightpink}{RGB}{255, 230, 235}

\usepackage{subcaption}
\usepackage{minted}
\definecolor{graybg}{RGB}{245, 245, 245}

\usepackage{tikz}
\usetikzlibrary{shapes.geometric}
\usetikzlibrary{calc}
\definecolor{michelinred}{RGB}{218,41,28}

\usepackage{hyperref}

\usepackage{listings}
\usepackage{enumitem}
\setlist{nosep}
\usepackage{titlesec}
\usepackage{amsfonts}

\def \globalscale {0.012}
\newcommand\michelin{
    \begin{tikzpicture}[y=1cm, x=1cm, yscale=\globalscale,xscale=\globalscale, every node/.append style={scale=\globalscale}, inner sep=0pt, outer sep=0pt]
      \path[draw=red, line width=0.6, line cap=round, line join=round, fill=none] 
        (12.118, 2.567) 
        .. controls (10.166, 2.567) and (9.384, 1.693) .. 
        (8.895, 0.275)
        .. controls (8.590, -0.900) and (8.467, -2.540) ..
        (9.948, -6.297)
        .. controls (7.902, -3.316) and (6.006, -1.826) ..
        (4.260, -1.826)
        .. controls (2.514, -1.826) and (1.402, -3.316) ..
        (1.402, -5.001)
        .. controls (1.402, -7.206) and (3.757, -8.502) ..
        (8.467, -8.890)
        -- (8.467, -9.657)
        .. controls (3.757, -10.063) and (1.402, -11.359) ..
        (1.402, -13.547)
        .. controls (1.402, -15.232) and (2.514, -16.722) ..
        (4.260, -16.722)
        .. controls (6.006, -16.722) and (7.902, -15.232) ..
        (9.948, -12.224)
        .. controls (8.467, -15.981) and (8.590, -17.621) ..
        (8.895, -18.796)
        .. controls (9.384, -20.214) and (10.166, -21.087) ..
        (12.118, -21.087)
        .. controls (14.070, -21.087) and (14.852, -20.214) ..
        (15.341, -18.796)
        .. controls (15.646, -17.621) and (15.769, -15.981) ..
        (14.288, -12.224)
        .. controls (16.334, -15.232) and (18.230, -16.722) ..
        (19.976, -16.722)
        .. controls (21.722, -16.722) and (22.834, -15.232) ..
        (22.834, -13.547)
        .. controls (22.834, -11.359) and (20.479, -10.063) ..
        (15.743, -9.657)
        -- (15.743, -8.890)
        .. controls (20.479, -8.502) and (22.834, -7.206) ..
        (22.834, -5.001)
        .. controls (22.834, -3.316) and (21.722, -1.826) ..
        (19.976, -1.826)
        .. controls (18.230, -1.826) and (16.334, -3.316) ..
        (14.288, -6.297)
        .. controls (15.769, -2.540) and (15.646, -0.900) ..
        (15.341, 0.275)
        .. controls (14.852, 1.693) and (14.070, 2.567) ..
        (12.118, 2.567)
        -- cycle;
    \end{tikzpicture}
}

\usepackage[dvipsnames]{xcolor}
\usepackage{xspace}

\lstdefinestyle{cmd}{
  basicstyle=\ttfamily\small,
  columns=fullflexible,
  breaklines=true,
  showstringspaces=false,
  frame=single,
  framerule=0.4pt,
  rulecolor=\color{black!25},
  backgroundcolor=\color{black!3},
  xleftmargin=0.6em,
  framexleftmargin=0.4em,
  aboveskip=0.6em,
  belowskip=0.6em
}

\newcommand{\ghas}{GitHub Actions\xspace}
\newcommand{\ghaworkflow}{GitHub Actions Workflow\xspace}
\newcommand{\ghaworkflows}{GitHub Actions Workflows\xspace}

\newcommand{\poutine}{{\fontfamily{cmss}\selectfont poutine}\xspace}
\newcommand{\ggshield}{{\fontfamily{cmss}\selectfont ggshield}\xspace}
\newcommand{\actionlint}{{\fontfamily{cmss}\selectfont actionlint}\xspace}
\newcommand{\zizmor}{{\fontfamily{cmss}\selectfont zizmor}\xspace}
\newcommand{\scorecard}{{\fontfamily{cmss}\selectfont scorecard}\xspace}
\newcommand{\frizbee}{{\fontfamily{cmss}\selectfont frizbee}\xspace}
\newcommand{\scharf}{{\fontfamily{cmss}\selectfont scharf}\xspace}
\newcommand{\pinny}{{\fontfamily{cmss}\selectfont pinny}\xspace}
\newcommand{\semgrep}{{\fontfamily{cmss}\selectfont semgrep}\xspace}
\newcommand{\nbscanner}{9\xspace}
\newcommand{\nbworkflows}{2,722\xspace}

\newcommand{\etal}{\textit{et al.}\xspace}
\newcommand{\nbopenrepos}{388\xspace}
\newcommand{\nbrules}{83\xspace}
\newcommand{\nbsample}{54\xspace}
\newcommand{\prctsample}{2\%\xspace}
\newcommand{\rev}[1]{\textcolor{black}{#1}}

\title{Unpacking Security Scanners for GitHub Actions Workflows}
\author{
    \IEEEauthorblockN{
        Madjda Fares,
        Yogya Gamage,
        Benoit Baudry
    } \\
   \IEEEauthorblockA{Université de Montréal, Montreal, Canada} \\
    {\{madjda.fares, yogya.gamage, benoit.baudry\}@umontreal.ca}
}

\begin{document}

\maketitle

\begin{abstract}

\ghas is a widely used platform to automate the build and deployment of software projects through configurable workflows. 
As the platform's popularity grows, it also  becomes a target of choice for  software supply chain attacks. 
These attacks exploit excessive permissions, ambiguous versions or the absence of artifact integrity checks to compromise the workflows. 
In response to these attacks, several security scanners have emerged to help developers harden their workflows.

In this paper, we perform the first systematic comparison of \nbscanner \ghaworkflows security scanners. 
\rev{We compare them regarding scope (which security weaknesses they target), detection capabilities (how many weaknesses they detect), accuracy (how well their findings match manually confirmed weaknesses), and performance (how long they take to scan a workflow).}
In order to compare the scanners on a common ground, we first establish a classification of 10 common security weaknesses that can be found in \ghaworkflows. Then, we run the scanners against a curated set of \nbworkflows workflows.

\rev{Our study reveals that static scanners for \ghaworkflows are very diverse, with both general purpose and focused scanners.}
More importantly, we provide evidence that these scanners implement fundamentally different analysis strategies, leading to major gaps regarding the nature and the number of reported security weaknesses. 
%\rev{A manual validation of \nbsample workflows further shows that scanners covering the same weaknesses can exhibit substantially different precision and recall.} 
Based on this empirical evidence we make actionable recommendations for developers to harden their \ghaworkflows.

\end{abstract}

\section{Introduction}

% Context: GHA, growing adoption, vector for supply chains attacks, emergence of scanning tools
\ghas (GHA) is a Continuous Integration/Continuous Deployment (CI/CD) platform that lets developers automate builds, tests, and deployments through event-driven workflows \cite{rostami2022use}. Since it is inherently integrated with GitHub and widely adopted across both open-source and enterprise projects, it has become a key element of the software supply chain \cite{enck2022challenges}. However, this popularity has also made GHA an attractive vector for software supply chain attacks \cite{nist2023cicd, 10179304}. These attacks exploit security weaknesses in \ghaworkflows, such as the possibility of injecting arbitrary code in the workflow, or tampering with a workflow's third-party dependencies  \cite{pan24,10.1145/3643991.3644899}.
For example, in March 2025, \texttt{tj-actions/changed-files}, a popular reusable action was maliciously updated compromising tens of thousands of dependent workflows that did not pin their reference to that action  \cite{tjactionscve}. %Such incidents are evidence that weaknesses in \ghaworkflows are currently exploited to carry out supply chain attacks. GHA

%Recent incidents, such as the supply chain attack on the GitHub Action \texttt{tj-actions/changed-files}, where a widely used repository of reusable GitHub Actions was maliciously updated compromising all unpinned references, demonstrate how a single workflow dependency can cascade into a supply chain attack entry point \cite{tjactionscve}.

To mitigate these risks, researchers in academia and industry have developed  scanners that statically analyze \ghaworkflows, searching for different security weaknesses.  For instance, GitGuardian’s \ggshield \cite{basak2023comparative} can detect unintentionally exposed secrets, and \actionlint can detect subtle syntax and type errors in the workflows. Yet, these various scanners vary in scope and often address complementary aspects of workflow security. As a result, it is challenging for developers to navigate this fragmented landscape and choose the most appropriate scanner for their needs.

%%%%%%%%%%%%%%%%%%%%%%%%%%%%%%%%%%%%%%%%%%%%%%%%%
Previous works have studied the usage of GitHub Actions in practice \cite{zhang2024discussion}, the challenges developers face when using GitHub Actions \cite{Cheenepalli2025sme}, and GitHub Action outdatedness \cite{decan2023outdatedness}. A few studies focus specifically on the security weaknesses of GitHub Actions \cite{khatami2024smells, koishybayev2024quantifying}.
%Khatami \etal \cite{khatami2024smells} investigate GitHub Action Workflow smells, and Koishybayev \etal \cite{koishybayev2024quantifying} study security issues in GitHub Actions. 
However, to our knowledge, there is no systematic analysis and comparison of the scope and consistency of \ghaworkflow security scanners.  

%%%%%%%%%%%%%%%%%%%%%%%%%%%%%%%%%%%%%%%%%%%%%%%%%
\rev{In this paper, we present the first systematic comparative analysis of locally executable static security scanners for \ghaworkflows.}
We begin with 30 open-source candidates collected from the research literature, GitHub repositories, and security blogs. In favor of reproducibility, we exclude scanners that require cloud execution. We also discard scanners that are not maintained. This way, we curate a set of \nbscanner open source, state-of-the-art workflow security scanners.
We classify distinct detection rules across the scanners into ten common security weaknesses, which we validate with the main maintainers of the scanners. This  makes it possible to compare scanners despite differences in rule names and levels of granularity. %The chosen weaknesses cover a broad range of issues, including injection, unpinned dependencies, excessive permissions, insecure triggers, and structural errors in workflows. 

To compare the capacities of the \nbscanner scanners, we curate a dataset of \nbworkflows real-world workflows. The workflows are collected from \nbopenrepos open-source repositories, hosted by organizations from software companies such as Microsoft, Google or NVIDIA. \rev{Our evaluation focuses on four aspects: \textit{Coverage}, by mapping each scanner's detection rules to the selected 10 weaknesses to determine which common security weaknesses are addressed;
\textit{Detection consistency}, by analyzing similarities and differences among the weaknesses reported by the different scanners;
\textit{Accuracy}, by comparing scanner findings against manually confirmed weaknesses in a sample of \nbsample workflows;
and \textit{Performance}, by measuring per-workflow execution time to evaluate the practicality of integrating these scanners into continuous integration pipelines.}

Our results show that no single scanner covers all identified security weaknesses. Some specialized scanners focus on a single weakness, while general-purpose scanners report broader sets of potential risks. \rev{Most importantly, we find significant differences in scanner design, leading not only to discrepancies in the number of reported weaknesses, but also to different accuracy trade-offs. Our manual validation shows that scanners targeting the same weakness classes can differ substantially in precision and recall, confirming that detection volume alone is not sufficient to characterize scanner behavior.}
\rev{In terms of performance, most scanners remain fast enough for practical integration into CI pipelines, although their execution times vary across scanners.}
These findings emphasize the importance of combining complementary scanners.

Our main contributions are as follows.
\begin{itemize}
    \item \rev{The first systematic comparison of \ghaworkflow scanners regarding their scope, detection consistency, and accuracy. This assessment helps developers understand both what scanners target and how reliably they detect supported weaknesses.}
    \item A curated set of \nbscanner state-of-the-art, open-source \ghas Security Workflow scanners 
    \item Guidelines for developers to integrate \ghaworkflow  scanners, depending on their security requirements
    \item \rev{An open science repository containing the workflow dataset, scanner outputs, analysis scripts, and the manually labeled workflow sample used for the accuracy evaluation: \url{https://github.com/sparkrew/github-actions-security}}
\end{itemize}

\section{Background}

In this section, we first introduce the key concepts and terminology of \ghaworkflow. We also discuss the software supply chain attack risks associated with this process.

% \todo{clarify that an 'action' can point to a file that is locally in the same repo as the workflow or to a remote file that is documented in the marketplace; clarify that an action can be another workflow (a yml file) or a bash script, JS script or docker file}

% \todo{can a shell script or JS script be on the marketplace without being 'wrapped' in a yml file?}

% \todo{key concepts: trigger, reused action, pinned action, secret, job, step, VMs, }

% \todo{CVE database and how actions and CVEs are related}

% \todo{what's the action marketplace; explain that each action's version is stored with a hash. The marketplace is a sort of package registry for actions}

% \todo{clarify that a workflow is located in a workflow directory}

\subsection{Terminology of GitHub Actions}
\ghas is a continuous integration and continuous delivery (CI/CD) platform within GitHub. It allows developers to define workflows as declarative automation pipelines written in YAML. These files are stored in a directory called \texttt{.github/\allowbreak workflows/}. The workflows execute in response to events (also called triggers) such as \texttt{push} or \texttt{pull\_request}, or manual invocations like \texttt{workflow\_dispatch} \cite{github2024understanding}. We introduce the terminology and concepts to implement a workflow.

A \textbf{\ghaworkflow} consists of one or more jobs, which are executed in parallel or sequentially. Each job runs in a fresh virtual machine (VM) or container (called a runner). Runners can be GitHub-hosted (ephemeral VMs automatically provisioned in the cloud) or self-hosted (custom machines configured by the user). 

A \textbf{job} is associated with a target environment via the \texttt{runs-on} key (e.g., \texttt{ubuntu-latest}), and can define environment variables (\texttt{env:}), execution context (e.g., \texttt{working-directory}), and secrets (injected at runtime from the repository's secret store), among others. A job inherits a default set of repository permissions through the auto-generated \texttt{GITHUB\_TOKEN}. It can be customized using the \texttt{permissions:} key to restrict or expand access to GitHub resources such as contents or issues. 

A \textbf{step} defined in a job is an individual command or invocation of reusable units called actions.  Workflow steps can use \texttt{run:} to execute shell commands or \texttt{uses:} to call an action. 

An \textbf{action} is a building block of automation, declared in a file called \texttt{action.yml} or \texttt{action.yaml}. It can be a JavaScript action, a Docker container action, or a composite action defined in YAML. Actions may be local to the same repository (e.g., uses: \texttt{./.github/actions/my-action)} or remote, hosted in another repository and referenced via \texttt{owner/repo@version}. Versions can be referenced by tags, branches, or commit SHAs, but the best practice regarding security is to pin to a specific commit SHA.
%In contrast, reusable workflows are full workflow files (\texttt{.yml}) that can also be called with \texttt{uses:} but live under \texttt{.github/workflows/}.

The \textbf{GitHub Marketplace} serves as a registry for publishing and discovering actions and reusable workflows. Each published action version corresponds to a specific commit hash, which enables reproducibility when workflow authors pin their references to that hash. Actions published to the Marketplace must declare an \texttt{action.yml} to be installable; standalone shell or JavaScript scripts cannot be published to the Marketplace without this wrapper.

%Vulnerabilities in popular actions are often assigned \emph{GitHub Security Advisories} (GHSAs) and Common Vulnerabilities and Exposures (CVEs), linking workflow security directly to the broader vulnerability disclosure ecosystem.

%This architecture reflects the modular and declarative nature of \ghas. Workflows are simple to author, version-controlled with the code, and support powerful automation via prebuilt or custom actions from the GitHub Marketplace.
% \vspace{-0.5cm}
\begin{figure}[ht]
\centering
\caption{An example of a \ghaworkflow. Excerpt from the \href{https://github.com/excalidraw/excalidraw/blob/master/.github/workflows/size-limit.yml}{size-limit} workflow from \href{https://github.com/excalidraw/excalidraw/tree/master}{excalidraw}.}
\label{lst:bundle-size}
\begin{minted}[
    bgcolor=graybg,
    linenos,
    numbersep=5pt,
    fontsize=\scriptsize,
    frame=single,
    baselinestretch=1.2,
    breaklines
]{yaml}
name: "Bundle Size check @excalidraw/excalidraw"
on:
  pull_request:
    branches:
      - master
jobs:
  size:
    runs-on: ubuntu-latest
    env:
      CI_JOB_NUMBER: 1
    steps:
      - name: Checkout repository
        uses: actions/checkout@v3
      ...
      - uses: andresz1/size-limit-action@v1
        with:
          github_token: ${{ secrets.GITHUB_TOKEN }}
          build_script: build:esm
          skip_step: install
          directory: packages/excalidraw
\end{minted}
\end{figure}

\autoref{lst:bundle-size} illustrates an example  workflow. 
It is triggered by a \texttt{pull\_request} event targeting the master branch. It contains a single job named \texttt{size}, which runs two different steps. 
Both steps reuse a third-party action. The \texttt{actions/checkout@v3} action is provided by the \texttt{actions} organization, managed by GitHub, and is used to  check out repository code.  \texttt{andresz1/size-limit-action@v1} is owned by \texttt{andresz1} and checks the size of the project bundle.
The job also uses a scoped \texttt{GITHUB\_TOKEN} as a credential and sets several configuration parameters (e.g., \texttt{build\_script, skip\_step, directory}) for the size-checking action.

\begin{figure*}[ht]
\centering
\includegraphics[width=\textwidth]{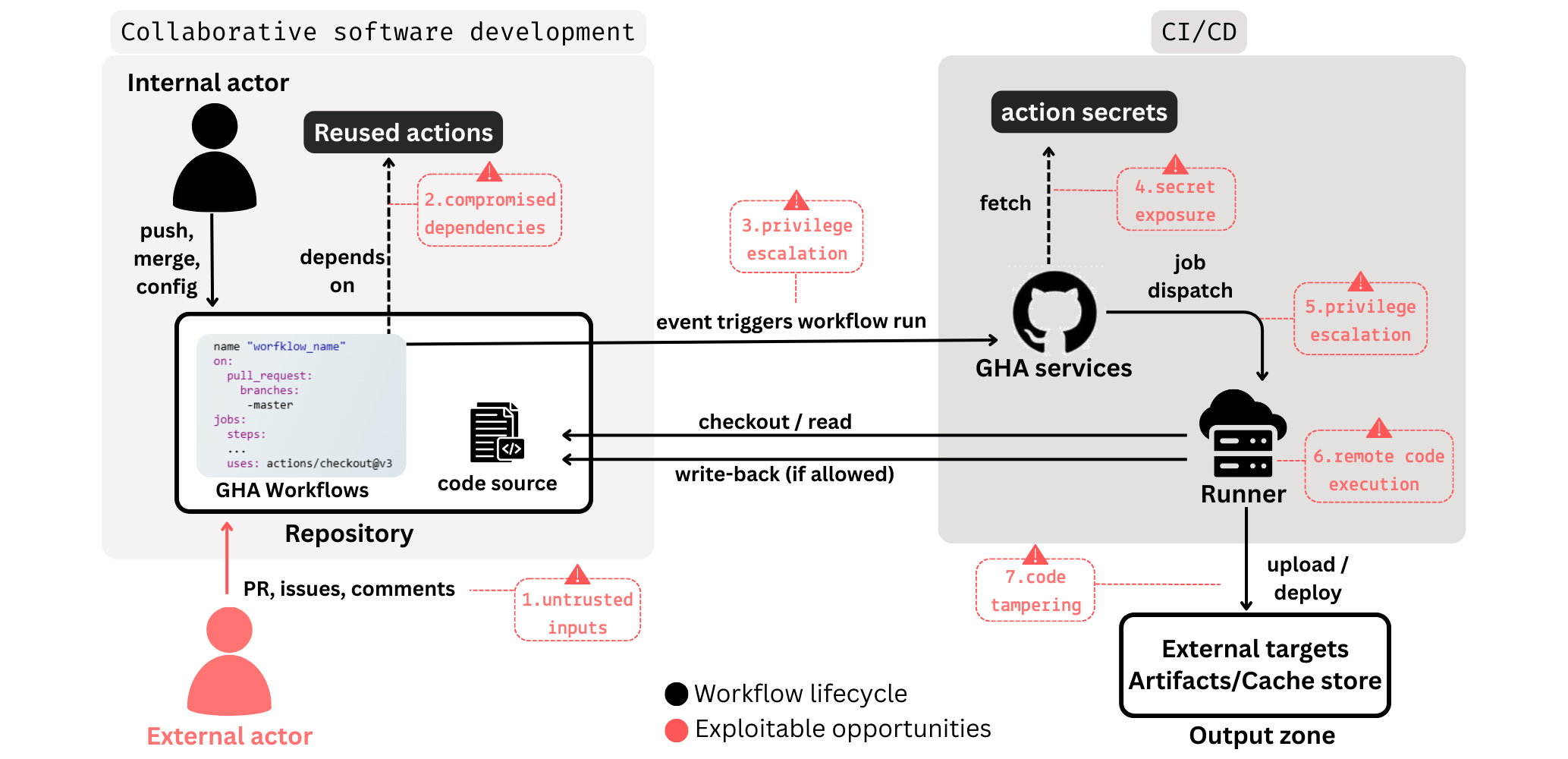}
\caption{GitHub Actions workflow lifecycle and key risk surfaces.}
\label{fig:GHA_diagram}
\end{figure*} 

\subsection{Risks in GitHub Actions Workflows}
 
In \autoref{fig:GHA_diagram}, we  summarize the regular process of continuous integration and continuous delivery (CI/CD) with \ghaworkflows (in black), as well as the associated risks, i.e. opportunities for software supply chain attacks that exploit the weaknesses in this process (in red). 

For regular CI/CD, software development teams build \ghaworkflows that automate build, dependency management, linting, testing, packaging, etc. 
The \ghaworkflows operate in the scope of a GitHub repository and are triggered every time the developers make a change, such as pushing and merging changes. 
A workflow typically reuses actions, e.g. to checkout the code (see \autoref{lst:bundle-size}) or to deploy a specific runtime for CI. 
When a workflow is triggered, \ghas services dispatch its jobs to a runner (e.g. a Docker container) and provide it with the necessary credentials and context (e.g., as environment variables). The runner checks out the  code from the repository, fetches referenced actions and dependencies, and executes the workflow steps. The runner reports logs and status to \ghas and may write results back to the repository when permissions allow. The runner can also produce outputs such as bundled artifacts and publish them on external registries.

In the following we discuss the key risks associated with this CI/CD process based on \ghaworkflows. First, external users who do not have write access on the repository can still interact with the repository, for example through issues or pull requests. These interactions can trigger \ghaworkflows and external users can pass malicious payloads (1) (e.g., \texttt{set a pull request title to '; curl http://evil.com/exfil | sh \#}). If workflows interpolate these event fields into shell commands, scripts, or environment variables, the payloads can enable injection during runner execution (6). Runner execution can  also be hijacked through reused actions; for example, in March 2026, the official \texttt{xygeni-action} \footnote{\url{https://www.stepsecurity.io/blog/xygeni-action-compromised-c2-reverse-shell-backdoor-injected-via-tag-poisoning}} was compromised through tag poisoning, causing repositories that referenced \texttt{@v5} to execute a C2 reverse shell on their runners. 
The code of third-party reused actions is fetched from third-party repositories, and it is possible for malicious actors to tamper with the referenced dependencies (2). Because these dependencies execute as part of the workflow run, a compromised upstream action can propagate to downstream repositories without changes in the workflow file itself. For example, a large supply chain attack was operated by tampering with the code of the popular third-party  \texttt{tj-actions/changed-files action} action  in March 2025.\footnote{\url{https://arstechnica.com/information-technology/2025/03/supply-chain-attack-exposing-credentials-affects-23k-users-of-tj-actions/}} 
In addition, if workflows rely on broad default permissions or do not explicitly scope \texttt{GITHUB\_TOKEN} access, a compromised step can abuse the token to escalate privileges or perform unintended write-back operations on the repository (3, 5). For example, the August 2024 Azure Karpenter Provider incident,\footnote{\url{https://www.stepsecurity.io/case-studies/azure-karpenter-provider}} where a pull-request-triggered workflow chain allowed a user-supplied ref from an untrusted fork to be checked out and executed in a privileged workflow context, leading to the exfiltration of privileged CI credentials. 
Additionally, although \ghas Services provide secrets only to authorized jobs at runtime, unsafe handling (e.g., logging or propagating secrets across steps) can lead to secret exposure (4).
Finally, workflows produce outputs such as artifacts with external targets. If these outputs are used without integrity checks, this creates output integrity and downstream impact risks (7), since tampered artifacts or cache entries can propagate to later stages or external systems.

\section{Workflow Security Scanners under Study}

In this section, we introduce our methodology to select state-of-the-art scanners that check for security weaknesses in \ghaworkflows through static analysis. In the rest of this paper, we refer to such tools as \emph{workflow security scanners}.

\subsection{Sourcing Open Source Workflow Security Scanners} 

The number and types of attacks against \ghaworkflows evolve fast and so does the landscape of workflow scanners. 
Consequently, we adopt an inclusive and exploratory approach to collect scanners. 
We concentrate on scanners that analyze workflows statically, as this is the state-of-the-art for scanning continuous integration workflows \cite{gu2023intrusion}. We only collect scanners that are open-source, to ensure accessibility, transparency, and reproducibility of our experiments.

To ensure a wide coverage, our initial collection process draws on different sources and discovery methods. 
First, we look for scanners in the security and software engineering literature. We search the proceedings of the following conferences: ICSE, ASE, FSE, ICSME, S\&P, USENIX security, and CCS. We focus on the past 5 years and use the following queries ``\textit{GitHub Action}'', ``\textit{continuous integration}'', ``\textit{workflow}''. We collect 39 papers and review each paper to determine if they propose a new scanner or if they use an existing scanner. We keep 14 papers, which refer to 11 different workflow security scanners, among which 4 propose new scanners.
Second, we search GitHub repositories tagged with the following topics: \texttt{github-actions security}, \texttt{supply-chain-security},  \texttt{devsecops}, \texttt{ci-cd}/\texttt{cicd}, \texttt{security-tools}, \texttt{scanner}, \texttt{dependency-security}. These tags often aggregate community-maintained scanners or best-practice implementations. We extract 10 workflow security scanners.
Third, we explore the grey literature. We go through 28 case studies, and blog articles published by organizations that are actively involved in software security, CI/CD tooling, and open-source security auditing. We identify 5 new scanners  through this process.  We also analyze the popular Reddit communities related to our study: \texttt{r/devops}, \texttt{r/github}, \texttt{r/cybersecurity}, \texttt{r/devsecops}, \texttt{r/pwnhub}, and extract 4 workflow security scanners.

This broad collection phase provides a list of 30 distinct scanners explicitly designed to secure \ghas workflows. Next, we analyze and curate this initial list for the purpose of our study.
%\href{https://github.com/Madjda32-del/github-actions-security/tree/main}{GitHub public repository}.

\subsection{Curated list of Workflow Security Scanners}

 In the following, we breakdown our curation process, to select a set of workflow security scanners that are amenable to sound and reproducible experiments.

\begin{table}[t]
\caption{Curated set of \ghaworkflows security scanners.}
\label{tab:opensourcescanners}
\centering
% \scriptsize
% \setlength{\tabcolsep}{3pt}
% \renewcommand{\arraystretch}{1.15}
\rowcolors{2}{gray!10}{white}

\begin{tabularx}{\columnwidth}{@{}l >{\ttfamily}p{0.95cm} p{1.55cm} p{0.8cm} Y@{}}
\toprule
Scanner & Commit & Collected & Lang. & Source \\
\midrule
\href{https://github.com/rhysd/actionlint}{\actionlint} & \href{https://github.com/rhysd/actionlint/commit/2ab3a12}{2ab3a12} & 2025-01-19 & Golang & \cite{zhang2024llmworkflows} \\
\href{https://github.com/stacklok/frizbee}{\frizbee} & \href{https://github.com/stacklok/frizbee/commit/bfdd688}{bfdd688} & 2025-05-31 & Golang & \href{https://stacklok.com/blog/new-frizbee-features-and-a-github-action-to-automate-pinning-actions-and-container-images}{Blog} \\
\href{https://github.com/GitGuardian/ggshield}{\ggshield} & \href{https://github.com/GitGuardian/ggshield/commit/a35fa49}{a35fa49} & 2025-06-04 & Python & \cite{basak2023comparative} \\
\href{https://github.com/koalalab-inc/pinny}{\pinny} & \href{https://github.com/koalalab-inc/pinny/commit/2e0d63a}{2e0d63a} & 2025-03-15 & Golang & \href{https://github.com/topics/supply-chain-security}{GitHub Topic} \\
\href{https://github.com/boostsecurityio/poutine}{\poutine} & \href{https://github.com/boostsecurityio/poutine/commit/68f289d}{68f289d} & 2025-06-02 & Golang & \href{https://github.com/topics/supply-chain-security}{GitHub Topic} \\
\href{https://github.com/cybrota/scharf}{\scharf} & \href{https://github.com/cybrota/scharf/commit/c47ec89}{c47ec89} & 2025-06-07 & Golang & \href{https://www.reddit.com/r/cybersecurity/comments/1jox7n8/scharf_an_opensource_scanner_to_identify_all/}{Reddit} \\
\href{https://github.com/ossf/scorecard}{\scorecard} & \href{https://github.com/ossf/scorecard/commit/ab2f6e9}{ab2f6e9} & 2025-05-30 & Golang & \cite{10.1145/3714464} \\
\href{https://github.com/returntocorp/semgrep}{\semgrep} & \href{https://github.com/returntocorp/semgrep/commit/7d387d2}{7d387d2} & 2025-06-08 & OCaml & \href{https://www.reddit.com/r/semgrep/comments/q1jrsh/protect_your_github_actions_with_semgrep//}{Reddit} \\
\href{https://github.com/zizmorcore/zizmor}{\zizmor} & \href{https://github.com/zizmorcore/zizmor/commit/af9b871}{af9b871} & 2025-06-09 & Rust & \href{https://www.wiz.io/blog/github-actions-security-guide}{Blog} \\
\bottomrule
\end{tabularx}
\end{table}

\textit{Workflow-focused target:} We focus on scanners that analyze \ghas workflow files (.yml or .yaml files). We exclude \href{https://github.com/latiotech/github-actions-log-checker}{GitHub Actions Log Checker} that scans the logs of the \ghas and \href{https://github.com/aquasecurity/trivy}{Trivy} that scans Container Images and the file system.

\textit{Reproducible execution:} Our study centers on scanners that we can configure and execute on our local machines. This allows us to precisely document the versions of the scanners and their configurations, for reproducibility purposes. In particular, we discard the tools that execute only on the GitHub runners or that depend on online scanners. \href{https://github.com/kondukto-io/kntrl}{Kntrl}, \href{https://github.com/step-security/harden-runner}{Harden-Runner}, \href{https://github.com/bullfrogsec/bullfrog}{Bullfrog}, \href{https://github.com/koalalab-inc/bolt}{BOLT} and \href{https://github.com/CycodeLabs/cimon-action}{Cimon} are not included in the benchmarking corpus.

\textit{Active projects:} To prioritize mature and actively maintained projects, we select scanners whose latest public commit was within the last six months. This guarantees relevance, and responsiveness to evolving \ghas features and security threats.We exclude \href{https://github.com/purs3lab/Argus}{Argus}, \href{https://github.com/Mobile-IoT-Security-Lab/GHAST}{GHAST}, \href{https://github.com/Ale0x78/GW-Checker}{GW-Checker}, and \href{https://github.com/sealuzh/cd-linter-artifacts}{CD-Linter}, as their last recorded commits date back more than 2 years.

\textit{Production ready:} We eliminate offensive simulation tools like \href{https://github.com/AdnaneKhan/Cacheract}{Cacheract}, which is designed to identify exploitable \ghas misconfigurations, and \href{https://github.com/Adnanekhan/Gato-X}{Gato-X}, which targets privilege escalation paths. These tools serve as red-team instruments rather than static analyzers, mainly intended to teach how GitHub Actions workflows can be exploited in practice.

\textit{Operational validation:} Finally, we execute the 11 remaining scanners on a random sample of 10 real \ghaworkflows. At this stage, we determine whether the scanners can run to completion locally and reproducibly.
After this step, we discard two scanners that cannot be run locally: CodeQL and \href{https://github.com/step-security/secure-repo}{Secure‑Repo}.
The CodeQL repository includes a set of rules to scan \ghaworkflows \footnote{\url{https://github.com/github/codeql/tree/main/actions}}. However, the execution of these rules in a local CLI, requires the  CodeQL CLI/bundle, which includes the YAML extractor. Yet, this extractor is not open-source, and we cannot run the CodeQL rules for scanning \ghaworkflows as part of our experiments. 
Regarding the Secure-Repo, its execution consists in invoking a cloud-based service that receives the workflow and runs the rules on a remote server. However, in order to have reproducible experiments, it is important that the rules and the rule engine run on our local machine. Consequently, we discard Secure‑Repo.

%https://github.blog/security/application-security/how-to-secure-your-github-actions-workflows-with-codeql/
%https://github.blog/changelog/2025-04-22-github-actions-workflow-security-analysis-with-codeql-is-now-generally-available/

\autoref{tab:opensourcescanners} provides the curated list of workflow security scanners that we use in this study. 
The table includes the name of the scanner, the specific commit ID used for our experiments, the date of this commit (as recorded during the collection period from June 3 to June 11, 2025), the primary programming language of the implementation, and the source through which we found this scanner. We also document the exact command we use to run each scanner, as part of our reproducibility package.\footnote{\url{https://github.com/sparkrew/github-actions-security/blob/main/scanners_output/command.md}}

% Each tool offers distinct detection rules and targets different classes of misconfigurations or vulnerabilities within \ghas workflows. We systematically analyse these different features in the next section.

\section{Methodology}

In this section, we present the methodology of our study. First, we explain how we build a classification of common weaknesses found in  \ghaworkflow security. Then, we present the dataset of \ghaworkflows against which we run the workflow security scanners. Finally, we introduce the research questions that guide our analysis.

\subsection{Classification of common weaknesses in \ghaworkflows}

While  security scanners all aim at analyzing \ghaworkflows, they implement rules with different names or different levels of granularity. %For example, \actionlint has 36 rules such as \texttt{untrusted-input}, which detects the unsafe usage of external input in shell commands, or \texttt{permissions-check}, which verifies whether workflows declare proper GitHub token scopes. Similarly, \zizmor provides 23 rules including \texttt{insecure-commands}, which flags risky shell commands, or \texttt{excessive-permissions}, which detects overly broad permission scopes in workflows. 
In order to support  a fair and interpretable comparison across scanners that use different naming conventions and detection granularity, we need to determine the common weaknesses that scanners search for. 
In order to classify the common security weaknesses, we proceed in four steps: manually analyze the documentation of the scanners, manually analyze the code of the scanning rules, cluster rules that detect similar weaknesses, validate our classification with the developers of the scanners.
%Each weakness represents a class of \ghaworkflow security risk. This level of abstraction supports a fair and interpretable comparison across scanners that use different naming conventions and detection granularity. 

\rev{First, two authors jointly review the documentation of each scanner to identify the rules it supports.
Because scanner documentation is not standardized, we first search for the keywords {\textit{rule}}, {\textit{pattern}}, and {\textit{check}}. This provides us with a list of candidate rule descriptions.}
%These keywords are search and grouping aids rather than the final classification method.}

\rev{Next, we cross-validate each candidate rule by inspecting the scanner source code.
We first search for the rule name in the implementation; when the name alone is insufficient, we use terms from the documentation (for example, {\textit{pin/unpinned}}, {\textit{permissions}}, {\textit{workflow}}, or {\textit{secret}}) to locate the corresponding detection logic and verify how the rule is implemented.
We also analyze the source code to identify additional rules that may not be documented, by following the program logic of the scanner from the main function, tracing method invocations, and identifying functions associated with rule implementation.
When the two authors disagree on whether a piece of code or a section of documentation corresponds to a single rule, they consult the third author for resolution.}
\rev{Consistent with the scope of our study, we retain only rules whose detection logic analyzes \ghaworkflows files under \texttt{.github/workflows/}.}

\rev{In total, we identify \nbrules rules across all scanners.
For each rule, the same two authors determine the type of vulnerability or misconfiguration it aims to detect, based on its documentation and code.
When we observe similarities in the goals of analysis rules across different scanners, we group them under the same common weakness.
These similarities are identified through a thorough analysis of documented behavior and implementation.
As before, in case of disagreement between the two authors when grouping rules under a weakness, they consult the third author.}

When clustering individual rules into weaknesses, we follow three principles: (1) we group rules that address the same security objective; (2) each rule is assigned to exactly one group; and (3) the weaknesses are defined and named in a way that rules from different scanners can be included, regardless of naming differences in different scanners.
This process leads to the definition of 10 common weaknesses.

Finally, to validate our interpretation of each scanner’s rule semantics and our mapping decisions, we contacted the lead maintainers of each scanner. We share (i) our classification of 10 common weaknesses, as well as (ii) the mapping between the rules of specific scanners and these common weaknesses. We ask the maintainers whether the mapping accurately reflects the intention of their scanner, and whether they agree that all of their rules can be expressed using our 10 common weaknesses. Out of 9 maintainers, 5 responded, 3 of them provided additional feedback, which we considered to adjust the description of 2 common weaknesses. All maintainers agreed with our mapping of rules to the common weaknesses.

The final classification of 10 common weaknesses is detailed in \autoref{sec:common} and forms the basis for our differential analysis of the 9 security scanners.

\subsection{Dataset of \ghaworkflows}

To compare the capacities of the workflow security scanners, we collect a dataset of workflows that are both representative of real-world usages and structurally diverse.
To build this benchmark, we mine  the repositories of the largest technology companies by revenue,\footnote{\url{https://en.wikipedia.org/wiki/List_of_largest_technology_companies_by_revenue}} which have at least one verified GitHub organization with public repositories.
% Our dataset includes workflows from the following companies and verified organizations: Amazon (\texttt{aws}, \texttt{awslabs}), Meta (\texttt{facebook}, \texttt{facebookresearch}, \texttt{facebookincubator}), Microsoft (\texttt{Azure}, \texttt{microsoft}), Tencent (\texttt{Tencent}), Apple (\texttt{apple}), Oracle (\texttt{oracle}), NVIDIA (\texttt{NVIDIA}), IBM (\texttt{IBM}), and Cisco (\texttt{cisco}).

For each company, we collect all their verified GitHub organizations, and we filter the repositories with a minimum of 2,000 GitHub stars, that contain at least one workflow file in the \texttt{.github/workflows/} directory and have received a commit within the 12 months preceding the collection date (Feb. 28, 2026).
For each repository, we collect all yaml files inside \texttt{.github/workflows/} directory. % and written in \texttt{.yml} or \texttt{.yaml} format. This ensures that all selected files are actual \ghas configurations executed by the platform. 
Our goal is to capture a broad range of CI/CD use cases reflected in real-world \ghaworkflows. 

% This includes automated tasks such as dependency management, build orchestration, testing, deployment, scheduled scanning, release automation, and linting. The diversity of use cases is reflected in the variety of triggers (\texttt{push, pull\_request, schedule, workflow\_dispatch}) and the number and types of third-party actions reused across the collected workflows.

%After collecting the workflows, we extract their content and normalize each file to improve consistency and parsing robustness. This process addresses common formatting issues such as inconsistent indentation and reserved keyword usage that may interfere with YAML parsers. We then parse the cleaned workflows to extract key structural elements.

\begin{table}[t]
\caption{Dataset breakdown by company (collected Feb.\ 27, 2026).}
\label{tab:dataset_breakdown_company}
\centering
\scriptsize
\setlength{\tabcolsep}{3pt}
\renewcommand{\arraystretch}{1.15}
\rowcolors{2}{white}{gray!10}

\begin{tabularx}{\columnwidth}{@{} l Y r r @{}}
\toprule
Company & GitHub verified organizations & \#Repos & \#Workflows \\
\midrule
Alphabet  & \href{https://github.com/google}{google} & 96 & 472 \\
Amazon    & \href{https://github.com/aws}{aws}, \href{https://github.com/awslabs}{awslabs} & 38 & 374 \\
Apple     & \href{https://github.com/apple}{apple} & 14 & 55 \\
Cisco     & \href{https://github.com/cisco}{cisco} & 2 & 2 \\
IBM       & \href{https://github.com/IBM}{IBM} & 3 & 33 \\
Meta      & \href{https://github.com/facebook}{facebook}, \href{https://github.com/facebookresearch}{facebookresearch}, \href{https://github.com/facebookincubator}{facebookincubator} & 71 & 358 \\
Microsoft & \href{https://github.com/Azure}{Azure}, \href{https://github.com/microsoft}{microsoft} & 130 & 1,083 \\
NVIDIA    & \href{https://github.com/NVIDIA}{NVIDIA} & 18 & 172 \\
Tencent   & \href{https://github.com/Tencent}{Tencent} & 16 & 173 \\
\midrule
\textbf{Total} (9) & \textbf{13} & \textbf{\nbopenrepos} & \textbf{\nbworkflows} \\
\bottomrule
\end{tabularx}
\end{table}

\begin{figure}[t]
\centering

  \includegraphics[width=\columnwidth]{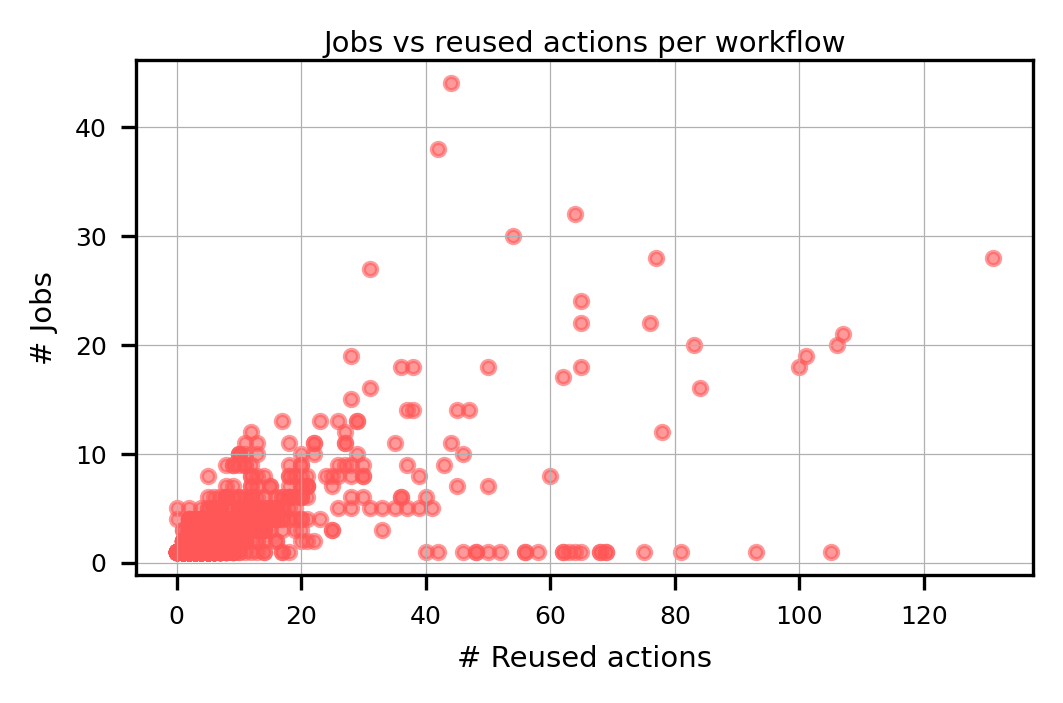}

\caption{Jobs vs. reused actions per workflow in our dataset (each point is one workflow)}
\label{fig:dataset_alpha_variation}
\end{figure}

The final dataset consists of \nbworkflows workflows collected from \nbopenrepos distinct open-source repositories.
\rev{This dataset intentionally focuses on repositories from large technology companies.
The goal is to obtain mature, high-impact, and structurally diverse workflows rather than to sample the entire GitHub ecosystem.
At the same time, these organizations typically follow stricter engineering and security practices than smaller or less security-conscious projects.
Consequently, the observed weakness distributions and scanner reporting behavior should not be interpreted as representative of all GitHub repositories.}

\autoref{tab:dataset_breakdown_company} reports the number of collected workflows per company, broken down by verified GitHub organization.
\autoref{fig:dataset_alpha_variation} relates two workflow complexity metrics across the full dataset: the number of reused actions (x-axis) and the number of jobs (y-axis), with each point representing one workflow. The plot shows a dense cluster of workflows with few jobs and few reused actions, alongside a sparse set of increasingly complex workflows that combine higher dependency usage and higher parallelism.
Median numbers of the reused actions and jobs are $3$ and $1$ respectively. Yet,  a long tail indicates increasingly complex workflows, with 193 workflows (7.1\%) that have more than 4 jobs and 1,086 workflows (39.9\%) that reuse more than 3 actions. 
The scatter plot also reveals clear outliers. For example, the workflow \href{https://github.com/facebook/rocksdb/blob/main/.github/workflows/pr-jobs.yml}{\texttt{pr-jobs.yml}} from \texttt{facebook/rocksdb} repo reuses $131$ actions. 
% This  high number of dependencies towards other actions is meant to \todo{what are the reused actions about?}. 
This outlier mainly stems from a multi-job ``mega-workflow'' that repeatedly calls repository-local (95) composite actions for CI setup, caching, and artifact/log collection, plus a smaller set of external utility actions (36).
%Yet, it also represents a large attack surface, as these reused workflows can be compromised or can change behavior over time. 
The workflow \href{https://github.com/microsoft/ebpf-for-windows/blob/main/.github/workflows/cicd.yml}{\texttt{cicd.yml}} is another example from the \texttt{microsoft/ebpf-for-windows} repository, which defines $44$  jobs, used to highly parallelize the continuous integration pipeline. 
%suggesting that workflows with many jobs often also rely on many reused actions.
These real-world workflows are diverse in complexity and functionality, and provide a good benchmark to exercise the different features of security scanners.

    % key structural characteristics of the workflows in our dataset, including distribution statistics for size, job count, trigger usage, and external action reuse. The median values suggest that most workflows are designed for simple, single-purpose automation tasks, while the upper bounds indicate the presence of more complex configurations with many jobs or dependencies.

%We illustrate these metrics with the workflow in \autoref{lst:bundle-size} from the \texttt{excalidraw/excalidraw} repository. It has 28 lines of YAML code, triggered by a \texttt{pull\_request} event. It defines a single job that reuses three external actions (\texttt{actions/checkout, actions/setup-node}, and \texttt{andresz1/ size-limit-action})

\label{sec:rqs}

\newcommand{\researchquestionone}{To what extent do state-of-the-art workflow security scanners search for the same workflow weaknesses?}

\newcommand{\researchquestiontwo}{To what extent do state-of-the-art workflow security scanners behave consistently regarding the detection of security weaknesses?}

\newcommand{\researchquestionaccuracy}{To what extent does the accuracy of weaknesses found vary across security scanners?}

\newcommand{\researchquestionthree}{What is the runtime performance of security scanners?}

\subsection{Research questions}

With our classification of common weaknesses in \ghaworkflows, and our dataset of \nbworkflows real-world workflows, we perform a systematic differential analysis of the security scanners. We articulate this analysis around four research questions that capture complementary aspects related to the scope of the scanners, their ability to report weaknesses, their accuracy, and their runtime performance.

\textbf{RQ1. [\textit{static}] \researchquestionone}

% \todo{add a short paragraph to explain (i) that we look at issues covered, type of input, ability to generate fixed workflows; (2) that we compare the number of issues they cover, as well as their flexibility in usage}
To answer this question, we use the classification of common  workflow security weaknesses presented in  (\autoref{sec:common}) to compare the functionalities of the different scanners. 
For each scanner, we map the individual rules to the common weaknesses. 
To complement this comparison, we also document the type of inputs they support and whether they can automatically generate fixed workflows for detected weaknesses. For that, we analyze their documentation and command-line options.

% First, we manually group the rules that have similar names across tools. Second, we check the rules' documentation and their implementation in the scanners' source code, in order to validate our interpretation. With this second step, we resolve ambiguities and make sure that we assign each rule to the appropriate common weakness. Third, we ask the lead developers of each scanner, whether they agree with our mapping from rules to common weaknesses. Five out of nine developers reply, and all them confirm that we have correctly mapped their scanner's rules to common weaknesses.

\textbf{RQ2. [\textit{dynamic}] \researchquestiontwo}

% \todo{add a paragraph to discuss how we collect data: run all the tools on our set of workflows + we perform differential analysis + add a link to a page that documents the exact config of each tool for this experiment}

% \todo{what raw data? (logs, reports, fixed workflow) and what do we count? 
% for example: actionlint generates a log for each workflow that it analyzes, which includes a list of issues, and what rule has detected the issue. We extract the total number of reported issues from these logs.
% }

To answer this question, we run the scanners against our dataset of real‑world \ghaworkflows. If a scanner supports customization through configuration files, command-line options, or execution modes, we run it with the documented default settings unless otherwise stated.
We document the exact command we use to run each scanner, as part of our reproducibility package.\footnote{\url{https://github.com/sparkrew/github-actions-security/blob/main/scanners_output/command.md}} For example, ~\autoref{lst:semgrep-command} illustrates the command we use to run \semgrep on workflow \texttt{airbnb\_javascript\_\_node.yml}.

\begin{lstlisting}[
  style=cmd,
  language=bash,
  caption={Example command to run Semgrep on a workflow file.},
  label={lst:semgrep-command},
  captionpos=b
]
$ semgrep --config p/github-actions .github/workflows/airbnb_javascript__node.yml
\end{lstlisting}

To  compare the findings across scanners, we post-process the raw outputs of each scanner execution, and map them to the common weaknesses. We implement custom rules and patterns to mine the raw output.  For each workflow, we store the list of reported weaknesses, along with their corresponding detection rules and line numbers. Eventually, we produce one consolidated output per workflow. We then  analyze  how consistently the scanners detect weaknesses across the same workflows.  
The scripts to process the scanners outputs are available in  our reproducibility package.\footnote{\url{https://github.com/sparkrew/github-actions-security/blob/main/scripts/results.ipynb}}
 %For example, \actionlint generates a plain‑text log per workflow listing all the issues it detects and the rules that detected them. This allows us to count and compare the number and types of findings across tools, as well as the number of distinct workflows in which these findings occur.

To understand why some scanners interpret the same workflow structure in different ways, we inspect the documentation and, when needed, the rule implementations to clarify each rule’s intended scope and triggering conditions. This qualitative assessment allows us to precisely explain discrepancies among scanners.

\rev{\textbf{RQ3. [\textit{accuracy}] \researchquestionaccuracy}}

\rev{The detection-volume analysis of RQ2 compares what scanners report, but it does not determine whether these reports correspond to actual weaknesses or whether some weaknesses are missed. To complement this analysis, we  manually label a sample of \nbsample workflows (\prctsample of the full dataset).}

\rev{We sample workflows to cover diverse structural and security-relevant characteristics. We operationalize this diversity using four three-level metrics: structural complexity based on jobs and steps, trigger risk, action dependency profile, and security surface. This sampling strategy allows us to include workflows that are likely to exercise different scanner rules.}

\rev{For each selected workflow, we manually inspect the YAML file and assign ground-truth labels using the weakness taxonomy presented in \autoref{sec:common}. If we manually determine that a workflow includes multiple weaknesses of the same type, then we include the number of weaknesses as part of the label. Two team members manually label the \nbsample workflows separately and cross-check their labels. In case of disagreement, a third team member is consulted to make the final decision.}

\rev{To answer RQ3, we compare these manual labels with the scanner labels obtained from the normalized outputs produced for RQ2. For each scanner and each workflow in the sample, we compare the number of reported findings with the number of manually confirmed findings for each weakness class. Then, we compute precision and recall for each scanner.}

\textbf{RQ4. [\textit{performance}] \researchquestionthree}

Security scanners are often integrated into continuous integration pipelines, where fast feedback is essential. A scanner that significantly slows down execution may discourage adoption or be disabled in production workflows. Therefore, evaluating runtime performance is critical to ensure these scanners provide meaningful findings without introducing unacceptable delays. To assess this, we measure performance by recording the time each scanner takes to analyze a single workflow file. We measure this time using the Unix \texttt{time} utility, and we record elapsed wall-clock time (\texttt{real}) as the primary performance metric, since it reflects end-to-end CI delay. In addition, we also record CPU time components (\texttt{user} and \texttt{sys}) to help interpret whether runtime is dominated by computation or by per-invocation overhead and I/O. 

Each measurement is done independently per workflow and repeated three times to account for natural variations in runtime, and we report the median over the three runs. Measurements are conducted across the full dataset to capture both average behavior and outliers. All experiments are run on a laptop equipped with an Intel(R) Core(TM) i7-10510U CPU @ base clock 1.80 GHz, turbo boost up to 2.30 GHz, 16GB RAM (15.8GB usable).

\section{Classification of common security weaknesses in \ghaworkflows }\label{sec:common}

\begin{table}
\footnotesize
\caption{A classification of ten security weaknesses  in \ghaworkflows and the corresponding CWEs }
\label{tab:classification_with_cwe}
\centering
\renewcommand{\arraystretch}{1.2}
\setlength{\tabcolsep}{6pt}
\rowcolors{2}{gray!10}{white}
\begin{tabular}{ll}
\toprule
Weakness & CWE  \\
\hline
Artifact Integrity Weakness (AIW) & \href{https://cwe.mitre.org/data/definitions/353.html}{CWE-353}, \href{https://cwe.mitre.org/data/definitions/494.html}{CWE-494}  \\
Control Flow Weakness (CFW) & \href{https://cwe.mitre.org/data/definitions/571.html}{CWE-571}  \\
Excessive Permission Weakness (EPW) & \href{https://cwe.mitre.org/data/definitions/250.html}{CWE-250}, \href{https://cwe.mitre.org/data/definitions/732.html}{CWE-732}  \\
GitHub Runner Compatibility Weakness (GRCW) & \href{https://cwe.mitre.org/data/definitions/477.html}{CWE-477}, \href{https://cwe.mitre.org/data/definitions/440.html}{CWE-440}  \\
Hardening Gap Weakness (HGW) & \href{https://cwe.mitre.org/data/definitions/223.html}{CWE-223}  \\
Injection Weakness (IW) & \href{https://cwe.mitre.org/data/definitions/20.html}{CWE-20}, \href{https://cwe.mitre.org/data/definitions/94.html}{CWE-94}  \\
Known Vulnerable Component Weakness (KVCW) & \href{https://cwe.mitre.org/data/definitions/1395.html}{CWE‑1395}   \\
Privileged Trigger Weakness (PTW) & \href{https://cwe.mitre.org/data/definitions/862.html}{CWE-862}   \\
Secrets Exposure Weakness (SEW) & \href{https://cwe.mitre.org/data/definitions/200.html}{CWE-200}, \href{https://cwe.mitre.org/data/definitions/522.html}{CWE-522}    \\
Unpinned Dependency Weakness (UDW) & \href{https://cwe.mitre.org/data/definitions/829.html}{CWE-829}   \\

\bottomrule
\end{tabular}
\end{table}

\autoref{tab:classification_with_cwe} summarizes the 10 common security weaknesses that can be detected by the security scanners. In order to ground our classification within an established security framework, we also list the CWEs that have the same scope as our common weaknesses. In the rest of the section, we outline the risks introduced by each weakness, and we illustrate it with examples taken from real projects.

\textbf{Artifact Integrity Weakness (AIW).}
\ghaworkflows that do not validate the integrity of artifacts or reused dependencies are vulnerable to supply chain attacks. Malicious actors can poison caches or inject backdoored artifacts into later stages of the pipeline, leading to compromised artifacts or lateral movement into trusted systems.

In \autoref{lst:aii}, the step downloads the artifact named \texttt{next-swc-binary} from the current run’s GitHub Actions artifact storage (via \texttt{actions/download-artifact@v4}); because no checksum or signature is verified after retrieval, a tampered artifact could be propagated into later stages, compromising the build or deployment.

\begin{center}
\begin{minipage}{\linewidth}
\begin{minted}[bgcolor=graybg, fontsize=\footnotesize, 
frame=single, 
breaklines, 
baselinestretch=1.2]{yaml}
- uses: actions/download-artifact@v4
  if: ${{ steps.docs-change.outputs.DOCS_CHANGE == 'nope' }}
  with:
    name: next-swc-binary
    path: packages/next-swc/native
\end{minted}
\captionof{listing}{Artifact Integrity Weakness (AIW) found in project \href{https://github.com/vercel/next.js/blob/4b66771895737170a06be242be1e5afc760142d4/.github/workflows/pull_request_stats.yml}{vercel/next.js}}
\label{lst:aii}
\end{minipage}
\end{center}
%Related CWE entries include \href{https://cwe.mitre.org/data/definitions/353.html}{CWE-353} (Missing Support for Integrity Check) and \href{https://cwe.mitre.org/data/definitions/494.html}{CWE-494} (Download of Code Without Integrity Check). Related GHSA advisories include \href{https://github.com/advisories/GHSA-cxww-7g56-2vh6}{GHSA-cxww-7g56-2vh6} (arbitrary file write via artifact path traversal).

\textbf{Control Flow Weakness (CFW).} 
In \ghaworkflows, developers use conditions to control when specific steps should be executed. However, attackers can exploit flawed conditions such as always being true (overly permissive) to bypass intended safeguards. This may lead to the unauthorized execution of sensitive jobs, including deployments or access to secrets.

\autoref{lst:cfi} illustrates a multi-line condition defined with YAML's folded scalar (\texttt{>}). This folded scalar is used to evaluate an expression that is defined on multiple lines. When the expression is boolean, the result of the evaluation is passed as a non-empty string (\texttt{"true"} or \texttt{"false"}). However, in GitHub Actions any non-empty string in \texttt{if:} evaluates to \texttt{true}, so the job always runs. 

\begin{center}
\begin{minipage}{\linewidth}
\begin{minted}[bgcolor=graybg, fontsize=\footnotesize, frame=single, breaklines, baselinestretch=1.2]{yaml}
if: >
  ${{ github.event.workflow_run.event == 'pull_request' &&
      github.event.workflow_run.conclusion == 'success' }}
\end{minted}
\captionof{listing}{Control Flow Issue (CFW) found in project \href{https://github.com/EbookFoundation/free-programming-books/blob/c4db26b05d56692e246f302e8f9d6478fb360f06/.github/workflows/comment-pr.yml}{EbookFoundation/free-programming-books}}
\label{lst:cfi}
\end{minipage}
\end{center}
%Related CWE entry: \href{https://cwe.mitre.org/data/definitions/571.html}{CWE-571} (Expression is Always True).

\textbf{Excessive Permission Weakness (EPW).}
Overprivileged workflows, tokens, jobs, or permissions, create opportunities for privilege escalation attacks. If a workflow does not explicitly restrict its \texttt{GITHUB\_TOKEN} permissions, an attacker who gains control of a job via injection or logic flaws can abuse broad permissions to push code, access secrets, or modify critical repository settings.% This turns otherwise limited exposure into a full compromise vector.

\autoref{lst:epi} shows an excerpt of a workflow that grants excessive write permissions to a job, which can let a compromised action or script modify repository contents or pull requests. If a compromised action or script runs in this job, the over-privileged token could be abused to make unauthorized code changes, tamper with pull requests, or facilitate further privilege escalation.

\begin{center}
\begin{minipage}{\linewidth}
\begin{minted}[bgcolor=graybg, fontsize=\footnotesize, frame=single, breaklines, baselinestretch=1.2]{yaml}
permissions:
  contents: write
  pull-requests: write
\end{minted}
\captionof{listing}{Excessive Permission Weakness (EPW) found in project \href{https://github.com/n8n-io/n8n/blob/46432da41b85b4f7593f4a7c93066c4fabaa4c4b/.github/workflows/release-create-pr.yml}{n8n-io/n8n}}
\label{lst:epi}
\end{minipage}
\end{center}
%Related CWE entries include \href{https://cwe.mitre.org/data/definitions/250.html}{CWE-250} (Execution with Unnecessary Privileges) and \href{https://cwe.mitre.org/data/definitions/732.html}{CWE-732} (Incorrect Permission Assignment for Critical Resource).

\textbf{GitHub Runner Compatibility Weakness (GRCW).}
    This weakness occurs when a workflow references an action, runtime, or workflow construct that cannot be reliably resolved, interpreted, or supported in the current \ghas environment. Common root causes include outdated major versions of actions that rely on deprecated language runtimes, as well as unrecognized or renamed workflow keys, contexts, or expressions that do not resolve under the \ghas specification. These incompatibilities, as well as workflow constructs whose behavior remains difficult to resolve or reason about statically, can cause jobs to fail or be skipped, unintentionally disabling required checks or security steps. Attackers can exploit this by introducing changes that downgrade actions, use deprecated constructs, or rely on analysis-obscuring patterns, which can cause guardrail jobs to error out or be skipped, or become harder for static security checks to analyze reliably. In repositories where such jobs are not strictly required or are marked with \texttt{continue-on-error}, this can lead to the removal of code scanning, permission tightening, or policy checks. This, in turn, increases the chance of subsequent abuse.

In \autoref{lst:gaci}, the workflow uses \texttt{actions/setup-node@v1}, an outdated action version that relies on a deprecated runtime and is no longer compatible with current GitHub-hosted runners.

\begin{center}
\begin{minipage}{\linewidth}
\begin{minted}[bgcolor=graybg, fontsize=\footnotesize, frame=single, breaklines, baselinestretch=1.2]{yaml}
- name: Use Node.js ${{ matrix.node-version }} 
  uses: actions/setup-node@v1  # deprecated; incompatible with current runners
\end{minted}
\captionof{listing}{GitHub Runner Compatibility Weakness (GRCW) found in project \href{https://github.com/Chalarangelo/30-seconds-of-code/blob/d560793a9aaebe7e5b4fc389ef15dd1621f313f0/.github/workflows/test.yml}{chalarangelo/30-seconds-of-code}}
\label{lst:gaci}
\end{minipage}
\end{center}
%Related CWE entries include \href{https://cwe.mitre.org/data/definitions/477.html}{CWE-477} (Use of Obsolete Function) and \href{https://cwe.mitre.org/data/definitions/440.html}{CWE-440} (Expected Behavior Violation).

\textbf{Hardening Gap Weakness (HGW).}
Workflows that lack integrated security scanning tools (e.g., SAST, dependency auditing, or secret scanning) create a hardening gap, giving attackers a broader surface to operate undetected. Without static analysis, dependency updates, or review enforcement, adversaries can introduce vulnerable code, exploit unpatched components, or suppress controls that would otherwise  expose their actions.

\autoref{lst:hgi} illustrates a standard build pipeline. The \texttt{pretest} step is a project-specific testing step. However, the workflow does not use or invoke any security testing scanner or scanner (e.g., static code analysis/SAST, dependency/advisory auditing, or secret scanning), creating a hardening gap where vulnerabilities can pass unnoticed.

\begin{center}
\begin{minipage}{\linewidth}
\begin{minted}[bgcolor=graybg, fontsize=\footnotesize, frame=single, breaklines, baselinestretch=1.2]{yaml}
steps:
  - uses: actions/checkout@v2
  - uses: ljharb/actions/node/install@main
  - run: npm run pretest
  # no code scanning or security testing step
\end{minted}
\captionof{listing}{Hardening Gap Weakness (HGW) found in project \href{https://github.com/airbnb/javascript/blob/11f986fdc7d6b4c80e396437e9c45c939362bdee/.github/workflows/node-pretest.yml}{airbnb/javascript}}
\label{lst:hgi}
\end{minipage}
\end{center}
%Related CWE entry includes \href{https://cwe.mitre.org/data/definitions/223.html}{CWE-223} (Omission of Security-Relevant Information), which often manifests as insufficient logging/monitoring.

\textbf{Injection Weakness (IW).}
Injection risks such as unsafe uses of untrusted inputs within shell or JavaScript contexts or environment variables in CI workflows allow attackers to execute arbitrary code by manipulating such inputs. They can inject commands through pull request metadata or workflow inputs, leading to code execution, data exfiltration, or alteration of pipeline behavior.

In \autoref{lst:ii}, the value of \texttt{inputs.run\_id}, potentially attacker-controlled when exposed via \texttt{workflow\_dispatch}, is expanded unquoted inside \texttt{run:}, enabling shell-metacharacter injection and arbitrary command execution. Mitigation includes restricting who can set the input, validating its format, or quoting it. 

\begin{center}
\begin{minipage}{\linewidth}
\begin{minted}[bgcolor=graybg, fontsize=\footnotesize, frame=single, breaklines, baselinestretch=1.2]{yaml}
run: |
    gh run watch ${{ inputs.run_id }} > /dev/null 2>&1
    gh run rerun ${{ inputs.run_id }} --failed
\end{minted}
\captionof{listing}{Injection Weakness (IW) found in project \href{https://github.com/facebook/react-native/blob/main/.github/workflows/retry-workflow.yml}{facebook/react-native}}
\label{lst:ii}
\end{minipage}
\end{center}
%Related CWE entries include \href{https://cwe.mitre.org/data/definitions/20.html}{CWE-20} (Improper Input Validation) and \href{https://cwe.mitre.org/data/definitions/94.html}{CWE-94} (Code Injection).

\textbf{Known Vulnerable Component Weakness (KVCW).}
Workflows that reuse and rely on components or actions with known vulnerabilities may expose the CI pipeline to existing exploits. An adversary can leverage these unpatched weaknesses to hijack execution, gain unauthorized access, or execute arbitrary code, especially in environments where updates are not regularly enforced.

In \autoref{lst:kvci}, the workflow uses \texttt{actions/download-artifact@v4}; versions 4.0.0 through 4.1.2 were vulnerable, and the issue was fixed in version 4.1.3.

\begin{center}
\begin{minipage}{\linewidth}
\begin{minted}[bgcolor=graybg, fontsize=\footnotesize, frame=single, breaklines, baselinestretch=1.2]{yaml}
- name: Download Windows binaries
  uses: actions/download-artifact@v4  # Known vulnerable version (ensure >= 4.1.3)
\end{minted}
\captionof{listing}{Known Vulnerable Component Weakness (KVCW) found in project \href{https://github.com/denoland/deno/blob/6e77e868967681c2d0f092c0f6047b2684978b61/.github/workflows/promote_to_release.yml}{denoland/deno}}
\label{lst:kvci}
\end{minipage}
\end{center}
%Related CWE entries include \href{https://cwe.mitre.org/data/definitions/1395.html}{CWE‑1395} (Dependency on Vulnerable Third‑Party Component). Related GHSA advisory: \href{https://github.com/advisories/GHSA-cxww-7g56-2vh6}{GHSA-cxww-7g56-2vh6} (Vulnerable action version).

\textbf{Privileged Trigger Weakness (PTW).} Some workflows are triggered by events that run in a trusted, base repository context, such as \texttt{pull\_request\_target} and \texttt{issue} or \texttt{comment}-driven automation. This means that users who do not have write access on the repository can still pass inputs (e.g., the name of an issue) to a workflow that runs with write-access privileges. A workflow that combines one of these triggers  with instructions that process untrusted data or code paths, turns into a serious risk. If the workflow checks out or executes PR-controlled code, or uses event fields in scripts, it can enable secret exfiltration or repository modification.

\begin{center}
\begin{minipage}{\linewidth}
\begin{minted}[bgcolor=graybg, fontsize=\footnotesize, frame=single, breaklines, baselinestretch=1.2]{yaml}
on: pull_request_target:
...
jobs: ...
  notify: ...
    steps: ...
      - uses: tsickert/discord-webhook@6dc739
               8f3f165f16dadc5666051c367efa1692f4
        with:
          webhook-url: ${{ secrets.COMPILER_DISCORD_WEBHOOK_URL }}
          embed-description: ${{ github.event.pull_request.body }}
\end{minted}
\captionof{listing}{Privileged Trigger Weakness (PTW) found in project \href{https://github.com/facebook/react/blob/main/.github/workflows/compiler_discord_notify.yml}{facebook/react}}
\label{lst:tmi}
\end{minipage}
\end{center}
%Related CWE entry: \href{https://cwe.mitre.org/data/definitions/862.html}{CWE-862} (Missing Authorization).

In \autoref{lst:tmi}, the workflow is triggered by \texttt{pull\_request\_target}, which runs in the base repository context and may access repository secrets and privileged tokens. In our example, the job uses a secret-backed Discord webhook while embedding attacker-controlled pull request metadata (for example the PR body). If authorization checks are missing or incorrect, an attacker can open a pull request to trigger privileged side effects and potentially leak secrets or perform unauthorized repository actions.

\textbf{Secrets Exposure Weakness (SEW).}
When sensitive credentials in workflows are not properly scoped or protected, such as hardcoded secrets, unnecessary exposure, or unsafe propagation across jobs, attackers can intercept them through log output, injected commands, or untrusted jobs. Leaked tokens, API keys, or credentials can be used to pivot into external infrastructure, spoof maintainers, or extract sensitive data, often without triggering immediate alarms in the CI environment.

\autoref{lst:sei} shows a workflow using \texttt{secrets: inherit}, which passes all repository secrets to a reusable workflow. This is unsafe if that workflow is external or untrusted because it gives external code access to sensitive credentials. 

\begin{center}
\begin{minipage}{\linewidth}
\begin{minted}[bgcolor=graybg, fontsize=\footnotesize, frame=single, breaklines, baselinestretch=1.2]{yaml}
jobs:
  build:
    uses: ./.github/workflows/build_reusable.yml
    secrets: inherit
\end{minted}
\captionof{listing}{Secrets Exposure Weakness (SEW) found in project \href{https://github.com/vercel/next.js/blob/4b66771895737170a06be242be1e5afc760142d4/.github/workflows/pull_request_stats.yml}{vercel/next.js}}
\label{lst:sei}
\end{minipage}
\end{center}
%Related CWE entries include \href{https://cwe.mitre.org/data/definitions/200.html}{CWE-200} (Exposure of Sensitive Information to an Unauthorized Actor) and \href{https://cwe.mitre.org/data/definitions/522.html}{CWE-522} (Insufficiently Protected Credentials).

\textbf{Unpinned Dependency Weakness (UDW).}
The absence of pinned versioning for actions or dependencies exposes workflows to substitution attacks, where an adversary introduces malicious updates into upstream components. By exploiting floating tags or unverified references, attackers can inject backdoors or tampered logic into workflows without any change in the workflow’s own source code, making the attack stealthy and persistent.

In \autoref{lst:udi}, the workflow references an action via the floating tag \texttt{@main}, but does not pin a specific version, which risks executing modified or malicious code if the referenced action is changed by a malicious actor. 

\begin{center}
\begin{minipage}{\linewidth}
\begin{minted}[bgcolor=graybg, fontsize=\footnotesize, frame=single, breaklines, baselinestretch=1.2]{yaml}
uses: freecodecamp/crowdin-action@main
\end{minted}
\captionof{listing}{Unpinned Dependency Weakness (UDW) found in project \href{https://github.com/freeCodeCamp/freeCodeCamp/blob/45c098d506c9c2e271b4a84e39ff8f7ec2f177a3/.github/workflows/crowdin-upload.curriculum.yml}{freeCodeCamp/freeCodeCamp}}
\label{lst:udi}
\end{minipage}
\end{center}
%Related CWE entries include \href{https://cwe.mitre.org/data/definitions/829.html}{CWE-829} (Inclusion of Functionality from Untrusted Control Sphere).

\section{Results}

In this section, we present the outcomes of our empirical comparison of \ghaworkflow security scanners, following the research questions introduced in \autoref{sec:rqs}.

\subsection{\researchquestionone}

\begin{table*}[t]
\centering
\caption{Weakness coverage and technical features per scanner. Weakness acronyms: AIW=Artifact Integrity Weakness, CFW=Control Flow Weakness, EPW=Excessive Permission Weakness, GRCW=GitHub Runner Compatibility Weakness, HGW=Hardening Gap Weakness, IW=Injection Weakness, KVCW=Known Vulnerable Component Weakness, PTW=Privileged Trigger Weakness, SEW=Secrets Exposure Weakness, UDW=Unpinned Dependency Weakness. Inputs:  the scanner expects the source code repository and / or the workflow files as input; the scanner's behavior can be configured. Fix WF: mark if the scanner generates a fixed workflow file.}
\label{tab:capability_coverage}
\footnotesize
\renewcommand{\arraystretch}{1.25}
\setlength{\tabcolsep}{3pt}
\rowcolors{2}{gray!10}{white}
\resizebox{\linewidth}{!}{
\begin{tabular}{|l|c|*{10}{>{\centering\arraybackslash}m{0.7cm}}|*{3}{>{\centering\arraybackslash}m{0.8cm}}|c|}
\hline
\textbf{Scanner Name} & \textbf{\#rules} & \multicolumn{10}{c|}{\textbf{Weaknesses (alphabetical order)}} & \multicolumn{3}{c|}{\textbf{Inputs}} & \textbf{\makecell{Fix\\WF}} \\
\cline{3-15}
& & AIW & CFW & EPW & GRCW & HGW & IW & KVCW & PTW & SEW & UDW 
& \makecell{Code\\Repo} & \makecell{WF\\ File} & \makecell{Config.\\File} & \\
\hline
\textbf{actionlint} & 36 &  & \cellcolor{lightgreen}\checkmark & \cellcolor{lightgreen}\checkmark & \cellcolor{lightgreen}\checkmark &  & \cellcolor{lightgreen}\checkmark &  &  & \cellcolor{lightgreen}\checkmark &  &  & \cellcolor{cyan!20}\checkmark & \cellcolor{cyan!20}\checkmark & \\
\textbf{frizbee} & 1 &  &  &  &  &  &  &  &  &  & \cellcolor{lightgreen}\checkmark & \cellcolor{cyan!20}\checkmark & \cellcolor{cyan!20}\checkmark & \cellcolor{cyan!20}\checkmark & \cellcolor{yellow!20}\checkmark \\
\textbf{ggshield} & 1 &  &  &  &  &  &  &  &  & \cellcolor{lightgreen}\checkmark &  & \cellcolor{cyan!20}\checkmark & \cellcolor{cyan!20}\checkmark & \cellcolor{cyan!20}\checkmark & \\
\textbf{pinny} & 1 &  &  &  &  &  &  &  &  &  & \cellcolor{lightgreen}\checkmark & \cellcolor{cyan!20}\checkmark &  &  & \cellcolor{yellow!20}\checkmark \\
\textbf{poutine} & 12 & \cellcolor{lightgreen}\checkmark & \cellcolor{lightgreen}\checkmark & \cellcolor{lightgreen}\checkmark &  &  & \cellcolor{lightgreen}\checkmark & \cellcolor{lightgreen}\checkmark & \cellcolor{lightgreen}\checkmark & \cellcolor{lightgreen}\checkmark &  & \cellcolor{cyan!20}\checkmark &  & \cellcolor{cyan!20}\checkmark & \\
\textbf{scharf} & 1 &  &  &  &  &  &  &  &  &  & \cellcolor{lightgreen}\checkmark & \cellcolor{cyan!20}\checkmark &  &  & \cellcolor{yellow!20}\checkmark \\
\textbf{scorecard} & 5 &  &  & \cellcolor{lightgreen}\checkmark &  & \cellcolor{lightgreen}\checkmark & \cellcolor{lightgreen}\checkmark &  & \cellcolor{lightgreen}\checkmark &  & \cellcolor{lightgreen}\checkmark & \cellcolor{cyan!20}\checkmark &  &  & \\
\textbf{semgrep} & 3 &  &  &  &  &  & \cellcolor{lightgreen}\checkmark &  & \cellcolor{lightgreen}\checkmark &  &  & \cellcolor{cyan!20}\checkmark & \cellcolor{cyan!20}\checkmark & \cellcolor{cyan!20}\checkmark & \cellcolor{yellow!20}\checkmark \\
\textbf{zizmor} & 23 & \cellcolor{lightgreen}\checkmark & \cellcolor{lightgreen}\checkmark & \cellcolor{lightgreen}\checkmark & \cellcolor{lightgreen}\checkmark & \cellcolor{lightgreen}\checkmark & \cellcolor{lightgreen}\checkmark & \cellcolor{lightgreen}\checkmark & \cellcolor{lightgreen}\checkmark & \cellcolor{lightgreen}\checkmark & \cellcolor{lightgreen}\checkmark & \cellcolor{cyan!20}\checkmark & \cellcolor{cyan!20}\checkmark & \cellcolor{cyan!20}\checkmark & \\
\hline
\end{tabular}
}
\end{table*}

\autoref{tab:capability_coverage}  summarizes the core features of the nine workflow security scanners under study. Each row corresponds to a specific scanner, while columns describe distinct dimensions that characterize their functionality. The columns are organized into five main sections. The \texttt{\#rules} column indicates the total number of distinct detection rules implemented by the scanner. We retrieve this number through manual inspection of each scanner’s documentation, rule catalogs, and source code. 
In the \texttt{Weaknesses} section, we map the scanners' rules to the common security weaknesses introduced in \autoref{sec:common}. A checkmark (\checkmark) indicates that the scanner includes at least one rule covering that weakness. 
The \texttt{Inputs} section captures the variety of options for the scanners: operate on standalone YAML workflow files (WF File); accept a full code repository including workflow directories (Code Repo); let users customize their behavior, through a configuration file (Config File). Configuration files allow users to tune the scanner for specific policies, set rule priorities, define exceptions, and control output or runtime behavior, thereby enabling more flexible and targeted workflow security auditing. %For instance \poutine supports a .poutine.yml as the config file.
The last column marks scanners that can auto-generate a fixed workflow file in addition to reporting weaknesses, such as, replacing floating action tags with immutable commit SHAs.
 % reflects the tools that not only detect issues but also generate automatic fixes, directly modifying the workflow files (e.g., replacing floating action versions with immutable commit SHAs). This demonstrates an issue detection capability beyond shallow YAML parsing, as these tools interpret workflow semantics to safely apply secure transformations. All other tools are limited to diagnosing issues, they highlight issues but leave remediation to the user.

To answer RQ1, we analyze the scanners in two groups. The first group includes the \emph{general-purpose scanners}: they aim for breadth, they implement many rules and cover several common weakness, often by combining lint-style checks with security and policy checks. The second group, the \emph{focused scanners}, includes the scanners that specialize in one or two common weaknesses, they typically have fewer rules, but these rules provide deeper or unique detection (and sometimes automated remediation).

\subsubsection{General-purpose scanners}

The general-purpose scanners include \actionlint, \zizmor, \poutine and \scorecard

\actionlint is a lightweight linter with the largest rule set in our study, and this breadth comes mainly from linting GitHub Actions semantics and authoring constraints.
\actionlint can detect five common  weaknesses in workflow files. Its strength is on advanced syntax-level analysis of YAML files to help developers enforce best practices. 
For example, it detects Control Flow Weaknesses (CFW) by locating logic flaws that could inadvertently trigger jobs under unsafe conditions. 
Most of the rules of \actionlint (28 out of 36) aim at the GitHub Runner Compatibility Weakness (GRCW). All these rules locate various syntactical weaknesses that make the workflow's YAML file incompatible with GitHub’s execution model. 
These rules are particularly important because they catch early configuration mistakes that could lead to insecure or failing automation pipelines.  
The configuration file of \actionlint allows developers to define self‑hosted runner labels, set allowed configuration variables, and ignore specific lint errors on a per‑path basis.  

% \textbf{\poutine and \zizmor are the most versatile scanners, with the broadest  coverage. }

\zizmor provides broad general-purpose workflow security scanning, with the largest coverage of the common security weaknesses.
\zizmor searches for all 10 weaknesses with 23 rules.
For example, it detects Secrets Exposure Weakness (SEW) by identifying credential fields or secret expansions within \texttt{container:} or \texttt{services:} definitions. It uses two complementary methods. First, it flags blanket pass-through configurations such as \verb|secrets: inherit| when invoking reusable workflows or actions. Second, it detects secret values expanded in \texttt{run:} steps or written to \texttt{\$GITHUB\_ENV}/\allowbreak \texttt{\$GITHUB\_OUTPUT}, where they may be exposed in logs.  \zizmor operates on both workflow files and repositories.  
Its \texttt{zizmor.yml} configuraiton file primarily supports the selection of specific rules or patterns. 
% \zizmor also has three rules to detect Injection Weaknesses (IW). It flags writes of untrusted data to \texttt{\$GITHUB\_ENV} or \texttt{\$GITHUB\_PATH}, detects the use of deprecated workflow commands such as \texttt{ACTIONS\_ALLOW\_UNSECURE\_COMMANDS}, \texttt{::set-env}, and \texttt{::add-path}, and matches untrusted \verb|${{...}}| template expansions inside \texttt{run:} or other execution contexts.

\poutine also covers many common weaknesses, but it is explicitly oriented toward supply-chain and third-party component risks.
\poutine, covers seven weaknesses with 12 rules and focuses on risks induced by third-party components.
For example, it detects Artifact Integrity Weaknesses (AIW) by checking for unverified or malicious repositories. 
%It inspects \texttt{actions/checkout} targets and PR contexts to see if the fetched reference originates from a fork or an external owner. When such an untrusted checkout is followed by execution primitives (for example, \texttt{run}, shell scripts, or build tools), the rule reports an AIW.
% It implements the Known Vulnerable Component Weakness (KVCW) by flagging usages of components with published CVEs. 
% It consults the Open Source Vulnerability (OSV) database to match the referenced action or build-component version against published advisories and raises a finding on a positive match.
\poutine requires a full repository as input. Its \texttt{.poutine.yml} configuration file offers fine-grained policy controls that allow selective rule enforcement, filtering by severity or vulnerability IDs, and defining allowed runners

\scorecard complements workflow-level scanning with repository-wide security posture checks, and it includes a small set of workflow-related checks that map to high-impact weaknesses.
\scorecard looks at the repository as a whole instead of only analyzing the workflow files. 
\scorecard currently searches for five types of weaknesses and places emphasis on repository-wide safeguards rather than workflow syntax.

Within the general-purpose group, \zizmor and \poutine have the widest coverage. \zizmor provides broad general-purpose security scanning, while \poutine prioritizes supply chain integrity with more flexibility.  \actionlint and \scorecard are slightly more focused. The key contrast is that \actionlint’s contribution is “many rules concentrated in fewer classes”, whereas \scorecard reaches a comparable coverage with a much smaller set of broad checks. 

\subsubsection{Focused scanners}

Focused scanners typically target one or two of the common weaknesses, with a small number of rules, but their detections can be uniquely deep (e.g., secret-finding heuristics) or uniquely actionable (e.g., automatic pinning fixes).

\semgrep is a focused scanner that can report potentially dangerous command injection patterns.
The 3 rules of \semgrep focus on Injection Weaknesses (IW) and Privileged Trigger Weaknesses (PTW). They detect deprecated workflow commands, dangerous shell constructs, unquoted variable expansions in \texttt{run:} steps, and unsafe combinations of \texttt{pull\_request\_target} with an explicit checkout of the PR head.
\semgrep uses flexible rule configuration files that define the security rules to run.
It accepts both individual workflow files and full repositories as input and can generate a new workflow file that fixes the detected weaknesses.

% Four tools, \frizbee, \ggshield, \pinny, and \scharf, each focus on one single weakness. 
\frizbee, \pinny and \scharf exclusively check for Unpinned Dependency Weaknesses (UDW). 
These three scanners also provide automated fixes by replacing floating action versions with pinned commit SHAs. This makes their analysis significantly more actionable than other scanners that only report weaknesses. Among them, \frizbee uniquely accepts both individual workflow YAML files and full repositories as input and its configuration file lets users exclude specific \ghas, branches, or container images from pinning.

\ggshield specializes in secrets detection (SEW). It uses a \texttt{.ggshield.yaml} (or global config) file to store authentication tokens and control secret‑scanning policies, such as token lifetimes, ignore patterns, and scan settings. 
\ggshield uses entropy scanning to detect secrets that are directly visible in the workflow and regular expressions to locate statements that might indirectly expose a secret. It supports scanning both workflow files and full repositories.

The focused group splits into two patterns. \frizbee, \pinny, and \scharf form a tight cluster with identical scope and similar output, \ggshield is similarly single-scope but in a different weakness family, and \semgrep is the only one that extends beyond a single class. The practical takeaway from their comparison is complementarity rather than competition.

\begin{tcolorbox}[boxrule=1pt,arc=.3em, left=4pt, right=4pt]
  \textbf{Answer to RQ1}: 
  % The landscape of workflow security scanners is characterized by general-purpose scanners on one hand and highly specialized scanners on the other hand. 
  \actionlint, \poutine, and \zizmor cover a variety of weaknesses, while \frizbee, \ggshield, \pinny, \scharf and \semgrep  targets one or two weaknesses. Some scanners are also capable of generating automated fixes. 
  This diversity highlights the importance of selecting scanners based on how they align with specific security requirements.
\end{tcolorbox}

\subsection{\researchquestiontwo}

\newcommand{\totcell}[1]{\cellcolor{gray!35}#1}

\begin{table*}[!t]
\centering
\caption{Number of security weaknesses detected by each workflow security scanner: number of detections is in \textbf{bold}, and the number of workflows where these weaknesses are detected is in regular font.}
\label{tab:detection_volume_matrix}
\footnotesize
\renewcommand{\arraystretch}{1.3}
\setlength{\tabcolsep}{5pt}

\resizebox{\linewidth}{!}{
\begin{tabular}{|l
|r r|r r|r r|r r|r r|r r|r r|r r|r r|r r|}
\hline
\textbf{Scanner} &
\multicolumn{2}{c|}{AIW} &
\multicolumn{2}{c|}{CFW} &
\multicolumn{2}{c|}{EPW} &
\multicolumn{2}{c|}{GRCW} &
\multicolumn{2}{c|}{HGW} &
\multicolumn{2}{c|}{IW} &
\multicolumn{2}{c|}{KVCW} &
\multicolumn{2}{c|}{PTW} &
\multicolumn{2}{c|}{SEW} &
\multicolumn{2}{c|}{UDW} \\
% \multicolumn{2}{c|}{Total} \\
\hline

\textbf{actionlint} &
\cellcolor{gray!15} &  &
\cellcolor{gray!15}\textbf{79} & 45 &
\cellcolor{gray!15}\textbf{24} & 24 &
\cellcolor{gray!15}\textbf{2,094} & 911 &
\cellcolor{gray!15} &  &
\cellcolor{gray!15}\textbf{31} & 21 &
\cellcolor{gray!15} &  &
\cellcolor{gray!15} &  &
\cellcolor{gray!15} \textbf{0} & 0 &
\cellcolor{gray!15} &   \\
% \totcell{\textbf{2228}} & \totcell{955} \\
\hline

\textbf{frizbee} &
\cellcolor{gray!15} &  &
\cellcolor{gray!15} &  &
\cellcolor{gray!15} &  &
\cellcolor{gray!15} &  &
\cellcolor{gray!15} &  &
\cellcolor{gray!15} &  &
\cellcolor{gray!15} &  &
\cellcolor{gray!15} &  &
\cellcolor{gray!15} &  & 
\cellcolor{gray!15}\textbf{8,020} & 1,645 \\
% \totcell{\textbf{8020}} & \totcell{1645} \\
\hline

\textbf{ggshield} &
\cellcolor{gray!15} &  &
\cellcolor{gray!15} &  &
\cellcolor{gray!15} &  &
\cellcolor{gray!15} &  &
\cellcolor{gray!15} &  &
\cellcolor{gray!15} &  &
\cellcolor{gray!15} &  &
\cellcolor{gray!15} &  &
\cellcolor{gray!15}\textbf{5} & 4 &
\cellcolor{gray!15} &   \\
% \totcell{\textbf{5}} & \totcell{4} \\
\hline

\textbf{pinny} &
\cellcolor{gray!15} &  &
\cellcolor{gray!15} &  &
\cellcolor{gray!15} &  &
\cellcolor{gray!15} &  &
\cellcolor{gray!15} &  &
\cellcolor{gray!15} &  &
\cellcolor{gray!15} &  &
\cellcolor{gray!15} &  &
\cellcolor{gray!15} &  &
\cellcolor{gray!15}\textbf{9,811} & 2,083 \\
% \totcell{\textbf{9811}} & \totcell{2083} \\
\hline

\textbf{poutine} &
\cellcolor{gray!15}\textbf{125} & 30 &
\cellcolor{gray!15}\textbf{6} & 6 &
\cellcolor{gray!15}\textbf{47} & 47 &
\cellcolor{gray!15} &  &
\cellcolor{gray!15} &  &
\cellcolor{gray!15}\textbf{107} & 61 &
\cellcolor{gray!15}\textbf{121} & 57 &
\cellcolor{gray!15}\textbf{201} & 112 &
\cellcolor{gray!15} \textbf{0} & 0 &
\cellcolor{gray!15} &   \\
% \totcell{\textbf{607}} & \totcell{298} \\
\hline

\textbf{scharf} &
\cellcolor{gray!15} &  &
\cellcolor{gray!15} &  &
\cellcolor{gray!15} &  &
\cellcolor{gray!15} &  &
\cellcolor{gray!15} &  &
\cellcolor{gray!15} &  &
\cellcolor{gray!15} &  &
\cellcolor{gray!15} &  &
\cellcolor{gray!15} &  &
\cellcolor{gray!15}\textbf{10,202} & 2,161  \\
% \totcell{\textbf{10202}} & \totcell{2161} \\
\hline

\textbf{scorecard} &
\cellcolor{gray!15} &  &
\cellcolor{gray!15} &  &
\cellcolor{gray!15}\textbf{2,349} & 1,600 &
\cellcolor{gray!15} &  &
\cellcolor{gray!15}\textbf{2,638} & 2,638 &
\cellcolor{gray!15}\textbf{5} & 5 &
\cellcolor{gray!15} &  &
\cellcolor{gray!15}\textbf{16} & 11 &
\cellcolor{gray!15} &  &
\cellcolor{gray!15}\textbf{11,347} & 2,178  \\
% \totcell{\textbf{6432}} & \totcell{2706} \\
\hline

\textbf{semgrep} &
\cellcolor{gray!15} &  &
\cellcolor{gray!15} &  &
\cellcolor{gray!15} &  &
\cellcolor{gray!15} &  &
\cellcolor{gray!15} &  &
\cellcolor{gray!15}\textbf{255} & 128 &
\cellcolor{gray!15} &  &
\cellcolor{gray!15}\textbf{15} & 10 &
\cellcolor{gray!15} &  &
\cellcolor{gray!15} &   \\
% \totcell{\textbf{270}} & \totcell{138} \\
\hline

\textbf{zizmor} &
\cellcolor{gray!15}\textbf{3,620} & 1,969 &
\cellcolor{gray!15} \textbf{0} & 0 &
\cellcolor{gray!15}\textbf{2,296} & 1,046 &
\cellcolor{gray!15} \textbf{0} & 0 &
\cellcolor{gray!15}\textbf{8} & 8 &
\cellcolor{gray!15}\textbf{1,778} & 432 &
\cellcolor{gray!15} \textbf{0} & 0 &
\cellcolor{gray!15}\textbf{196} & 184 &
\cellcolor{gray!15}\textbf{89} & 48 &
\cellcolor{gray!15}\textbf{2,003} & 959  \\
% \totcell{\textbf{9990}} & \totcell{2438} \\
\hline

% \textbf{Total} &
% \totcell{\textbf{3596}} & \totcell{1977} &
% \totcell{\textbf{85}} & \totcell{45} &
% \totcell{\textbf{3967}} & \totcell{1711} &
% \totcell{\textbf{2094}} & \totcell{911} &
% \totcell{\textbf{2646}} & \totcell{2638} &
% \totcell{\textbf{1316}} & \totcell{466} &
% \totcell{\textbf{121}} & \totcell{57} &
% \totcell{\textbf{411}} & \totcell{292} &
% \totcell{\textbf{94}} & \totcell{52} &
% \totcell{\textbf{24464}} & \totcell{2264} &
% \totcell{\textbf{38794}} & \totcell{2722} \\
% \hline

\end{tabular}
}
\end{table*}

\autoref{tab:detection_volume_matrix} presents the number of weaknesses detected by each scanner, across our dataset of  \nbworkflows \ghaworkflows. 
A value of 0 indicates that the scanner includes at least one rule for that weakness but, in practice, does not detect any weakness in the analyzed workflows. The number in bold font indicates the total count of findings  reported by the scanner, while the number in regular font indicates how many distinct workflows contain at least one such finding (e.g., \actionlint reports 79 control flow weaknesses (CFWs) across 45 workflows).

\subsubsection{General-purpose scanners}

The four general-purpose scanners identify similar types of common weakness, but report different numbers, depending on their scope and analysis strategies. In the following we highlight the  design choices that explain the disparities among the number of findings.

\poutine and \zizmor essentially search for the same common weaknesses. Yet, the two scanners implement different strategies to detect the weaknesses, which explains many of their differences. \rev{In particular, \poutine implements more targeted heuristics that focus on specific high-risk patterns.} Meanwhile, \zizmor implements rules that have a broader scope, similar to a linter, reporting a blend of vulnerabilities, anti-patterns, and formatting issues. These differences are particularly visible in the case of AIW, EPW, IW, and SEW.

\poutine and \zizmor respectively detect 125 and  3,620 Artifact Integrity Weaknesses (AIW).  \rev{\poutine flags checked-out content that can later be executed, including local actions, whereas \zizmor also reports risks around uploaded artifacts, credential persistence, and cache reuse, which provide broader contextual signals for possible integrity issues.}
For Excessive Permission Weaknesses (EPW), \poutine checks permissions only in specific, high‑risk situations, such as workflows triggered by events that can receive external inputs, when \zizmor applies broader rules by also checking job‑level settings and reporting any potentially unnecessary or overly broad permissions.
\poutine reports 107 Injection Weaknesses (IW) in 61 workflows, searching for scripts that are run without proper checks and unsafe debug settings. \zizmor's 1,778 IWs are induced by a broader injection model that flags any \verb|${{...}}| expansion in execution contexts and treats writes to \texttt{GITHUB\_ENV}/\texttt{GITHUB\_PATH} as sinks. 
\poutine's detection of Secrets Exposure Weaknesses (SEW) focuses on jobs that might implicitly expose  configured secrets or run in privileged environments where secrets could leak, which never occurs in our dataset. Meanwhile \zizmor reports 89 SEWs as it also flags risky secret handling practices, such as passing all secrets to reusable workflows or writing secrets into environment files.
Finally, \zizmor detects 2,003 Unpinned Dependency Weaknesses (UDW), all third-party actions and container images that are not pinned.

Only in the case of  Privileged Trigger Weaknesses (PTW), \poutine reports slightly more weaknesses than \zizmor (201 and 196 respectively). Yet, these  weaknesses are of different nature as \poutine  reports a PTW when it can match a concrete unsafe combination, such as a privileged or externally influenced trigger paired with risky follow-up behavior, while \zizmor detects the presence of privileged triggers but does not analyze follow-up behavior that processes external inputs.
 
\poutine detects 6 Control Flow Weaknesses (CFW) with its rule that identify expressions that are always true, and 121  Known Vulnerable Component Weaknesses (KVCW) by recognizing references to known vulnerable actions or versions. Meanwhile, \zizmor does not detect any of these weaknesses despite their documentation mentioning these weaknesses.

\actionlint is a general-purpose workflow security scanner, and the only scanner to report GitHub Runner Compatibility Weaknesses (GRCW). These weaknesses relate to syntactic features in workflow, which might be misinterpreted by the  Github actions runner. While these incompatibilities are not vulnerabilities per se, they can be exploited by malicious actors to bypass certain guardrails in the runner. This unique feature makes \actionlint a valuable scanner for comprehensive workflow audits. It also detects 79 CFW, 24 EPW, and 31 IW with rules essentially matching invalid syntax and elevated top-level scopes such as \texttt{contents: write} and \texttt{pull-requests: write} or unsafe shell commands and old workflow commands, e.g. \texttt{::set-output}. \rev{However, these rules  capture  syntactic patterns rather than security-specific patterns targeted by scanners such as \poutine.}

\scorecard is the only general-purpose scanner under study that does not only analyze workflows, but can also analyze source code or the health of a whole repository. In that sense, it acts more as a linter for \ghaworkflows than a security-oriented scanner. It detects significantly more EPWs (2,349) and HGWs (2,638) than the other scanners as it systematically reports anti-patterns such as the omission of a top-level \texttt{permissions:} declaration or the absence of a SAST tool, without further analysis.
\scorecard reports 5 Injection Weaknesses, flagging conditional script execution based on untrusted input from commit messages, a pattern that is also detected by other scanners. \scorecard reports PTWs when privileged triggers are used in ways that clearly enable untrusted code execution, e.g. \texttt{pull\_request\_target} with a checkout of the incoming pull request. 
Finally, \scorecard reports the highest number of Unpinned Dependency Weaknesses as it includes tooling dependencies such as package installers (11,347).

Our  analysis of the general-purpose  scanners reveals that  \rev{\poutine is comparatively the most conservative in detection volume, reporting targeted findings for a smaller set of patterns. \zizmor increases coverage through broader rules, but these findings require more triage to distinguish confirmed weaknesses from lower-relevance cases.} \actionlint is especially valuable for GitHub Runner compatibility issues, where it provides coverage that the other scanners largely do not. \scorecard has a holistic approach for evaluating a repository's health, but its \ghaworkflows abilities are less informative than workflow-specific scanners.

\subsubsection{Focused scanners}

Focused scanners contribute narrow but high-impact signals, either by specializing in one weakness family with strong remediation, or by applying a dedicated detection strategy.

\frizbee, \pinny, and \scharf focus specifically on scanning workflows for Unpinned Dependency Weaknesses (UDW). They detect 8,020, 9,811 and 10,202 weaknesses respectively. Our analysis of each scanner's rules reveals that \frizbee reports the lowest number because it discards reused actions that refer to the \texttt{main} or \texttt{master} branch. \pinny performs a comprehensive analysis of all dependencies, and finds 10,470 UDWs. However, a close analysis of these results, revealed that \pinny reports some dependencies that are actually pinned. Removing these pinned dependencies from \pinny's reported unpinned dependencies,  9,811 UDWs remain. \scharf  reports the largest number of UDWs, but it detects weaknesses even in lines that are commented out. \rev{Among the three, \frizbee is the most conservative in reporting, as its lower count reflects a deliberate exclusion policy, while \pinny and \scharf report more UDW cases.}

\ggshield focuses on  Secrets Exposure Weaknesses and detects 5 of them in our dataset. This scanner searches for hardcoded secrets, such as tokens or keys, using high-entropy detection \cite{gitguardian2025genericentropy}, rather than broadly analyzing workflow logic for injection patterns. This is a useful complement of workflow-logic scanners, such as \zizmor and \poutine, which primarily flag risky secret-handling practices rather than identifying exposed secret material itself. 

Similar to \scorecard, \semgrep is a scanner that is not specific to \ghaworkflows and can analyze more than 30 languages, from \texttt{yaml} to \texttt{solidity}. Its rules are conservative and pattern-based, which detect relatively few weaknesses. For IW, it flags risky command execution patterns and workflow-command misuse (including deprecated command usage and unsafe shell constructs) in \texttt{run:} contexts.
For PTW, it targets the canonical high-risk combination where \texttt{pull\_request\_target} is used together with a checkout of the pull request’s code.
It finds the exact same weaknesses as  \scorecard except in one case, when the checkout target is encoded as a conditional GitHub expression rather than a direct pull request head reference, indicating a rule-level recall gap rather than a conceptual mismatch.

Overall, the focused scanners differ primarily in their detection style. \rev{\frizbee is stricter by design and excludes \texttt{main}/\texttt{master} references, whereas \pinny and \scharf use broader  strategies that can require additional validation, such as already pinned or commented actions.} \ggshield is specialized in detecting secret material, making it complementary to logic-oriented scanners. \semgrep provides focused coverage, particularly for injection weaknesses, but is less broad for privileged-trigger weaknesses.

%The number of distinct workflows generally follows the same trend as the overall detection counts, meaning that tools reporting more findings also tend to flag a larger set of workflows. However, relying only on raw detection numbers is not sufficient for evaluating precision, as the number of distinct workflows affected provides essential context for understanding how concentrated or widespread these detections are.

%Overall, this analysis highlights that  each scanner implements its own interpretation of the security issues. Some issues, such as Injection Issue (II) and Secrets Exposure Issue (SEW), still need more mature and consistent approaches, given the wide variation in how potential problems are defined and detected. \poutine provides broad issue detection coverage, while \actionlint is valuable for its strong GitHub runner compatibility issue detections, and \frizbee excels at detecting unpinned dependency issues (UDW). This diversity shows that no single tool currently offers complete coverage, and combining complementary tools remains necessary for a more reliable assessment.

\begin{tcolorbox}[boxrule=1pt,arc=.3em, left=4pt, right=4pt]
  \textbf{Answer to RQ2}: Significant variations in the number of reported weaknesses reveal essential differences in the design of static security workflow scanners. \rev{On one hand, some scanners use targeted security rules and report relatively few weaknesses. On the other hand, other scanners apply broader, linter-like rules and report substantially more findings.}
\rev{Among the general-purpose scanners, \poutine is the most conservative in detection volume, while \zizmor applies broader rules that report a wider range of security and configuration issues. \actionlint provides strong structural checks, and \frizbee provides focused support for version pinning. Combining complementary scanners is necessary for reliable \ghaworkflows security assessment.}
\end{tcolorbox}

\subsection{\researchquestionaccuracy}

\begin{table}[!htbp]
\centering
\caption{\rev{Manual ground-truth weaknesses in the \nbsample-workflow  sample. The table reports both the total number of manually confirmed detections and the number of workflows containing at least one weakness of each type.}}
\label{tab:rq3_ground_truth}
\footnotesize
\renewcommand{\arraystretch}{1.15}
\rowcolors{2}{gray!10}{white}
\begin{tabular}{|l|r|r|}
\hline
\textbf{Weakness} & \textbf{Confirmed detections} & \textbf{Workflows} \\
\hline
AIW  & 7   & 6  \\
CFW  & 0   & 0  \\
EPW  & 35  & 31 \\
GRCW & 18  & 11 \\
HGW  & 29  & 29 \\
IW   & 14  & 5  \\
KVCW & 8   & 4  \\
PTW  & 3   & 3  \\
SEW  & 4   & 2  \\
UDW  & 330 & 45 \\
\hline
\textbf{Any weakness} & 448 & 50 \\
\hline
\end{tabular}
\end{table}

\rev{Our manual analysis of \nbsample \ghaworkflows reveals 448 weaknesses across 9 different types.
\autoref{tab:rq3_ground_truth} summarizes our findings. UDW dominates the ground-truth with 330 manually confirmed detections across 45 workflows. Overall, 50 of the \nbsample workflows contain at least one manually confirmed weakness. EPW, GRCW, HGW and IW are also frequent, appearing in 31, 11, 29 and 5 workflows respectively. In contrast, AIW, KVCW, PTW and SEW are rare in the sample, and CFW does not appear in any manually confirmed label.}% This distribution is important when interpreting scanner accuracy: scanners targeting common weaknesses, such as UDW or EPW, can be evaluated on many positive cases, whereas scanners targeting rare or absent weaknesses have fewer opportunities to obtain true positives.}

\begin{figure}[!htbp]
\centering
\includegraphics[width=.95\columnwidth]{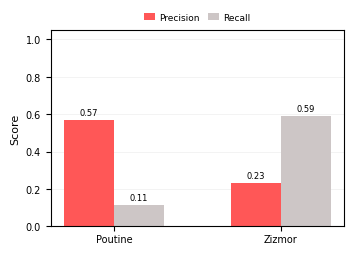}
\caption{\rev{Precision and recall of \poutine and \zizmor on the weakness classes covered by both: AIW, CFW, EPW, IW, KVCW, PTW, and SEW.}}
\label{fig:rq3_poutine_zizmor}
\end{figure}

\rev{\autoref{fig:rq3_poutine_zizmor} shows the precision and recall for the general-purpose scanners \poutine and \zizmor. We compare their accuracy on the weakness classes that are covered by both: AIW, CFW, EPW, IW, KVCW, PTW, and SEW. This comparison refines the detection-volume results of RQ2. In RQ2, \poutine reports far fewer findings than \zizmor on most shared weakness classes, which reflects its more targeted detection strategy. The ground-truth validation confirms this difference: \poutine produces fewer false positives than \zizmor (precision of 0.57 and 0.23, respectively), but also misses more true positives (recall of 0.11 and 0.59, respectively).}

\begin{figure}[!htbp]
\centering
\includegraphics[width=.95\columnwidth]{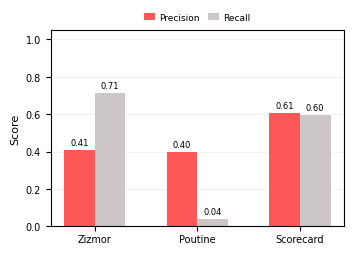}
\caption{\rev{Precision and recall of \zizmor, \poutine, and \scorecard on their common weakness scope: EPW, IW, and PTW.}}
\label{fig:rq3_common_scope}
\end{figure}

\rev{\autoref{fig:rq3_common_scope} restricts the comparison to the three weakness classes jointly covered by \zizmor, \poutine, and \scorecard: EPW, IW, and PTW. This controlled comparison shows that scanners with overlapping scope still differ substantially in accuracy. \zizmor obtains the highest recall, 0.71, which is consistent with its broader reporting strategy discussed in RQ2. \scorecard provides the most balanced behavior on this shared scope, with precision of 0.61 and recall of 0.60. \poutine has precision comparable to \zizmor, 0.40, with lower recall, 0.04. This suggests a more restrictive reporting behavior, with fewer false positives than \zizmor but also many missed manually confirmed EPW, IW, and PTW cases in the sample. These results show that weakness coverage alone does not determine scanner usefulness: even when scanners target the same weakness classes, their reporting rules lead to different accuracy tradeoffs.}

\begin{figure}[!htbp]
\centering
\includegraphics[width=.95\columnwidth]{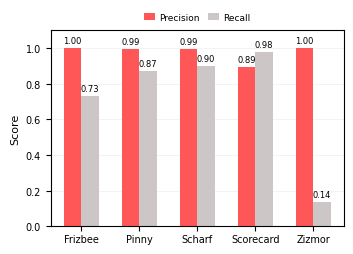}
\caption{\rev{Precision and recall of scanners that report Unpinned Dependency Weaknesses (UDW).}}
\label{fig:rq3_udw}
\end{figure}

\rev{\autoref{fig:rq3_udw} focuses on the 330 Unpinned Dependency Weaknesses in the ground-truth. Five scanners report UDW: \frizbee, \pinny, \scharf, \scorecard, and \zizmor. While scanners search for the same specific weakness, they still differ in accuracy. \frizbee reaches perfect precision, 1.00, but lower recall, 0.73, confirming that its stricter reporting strategy avoids false positives. \pinny and \scharf provide the strongest balance, with precision of 0.99 and recall of 0.87 and 0.90 respectively. \scorecard obtains the highest recall, 0.98, but lower precision, 0.89, which is consistent with its broader interpretation of unpinned dependencies in RQ2. Finally, \zizmor also reaches perfect precision on UDW, with a 0.14 recall, showing that it reports only few of the manually confirmed UDW cases.}

\rev{The \actionlint scanner reaches a precision of 0.44 and a recall of 0.25 on its supported weakness classes. It successfully detects all GRCWs, but misses EPW, IW, and SEW cases. This confirms that \actionlint is most useful as a structural and compatibility-oriented scanner. \semgrep searches for IW and PTW, with a precision and recall of 0.73 and 0.57 for IW and 1.0 and 0.33 for PTW. \semgrep identifies a meaningful portion of injection weaknesses with relatively few false positives, and its narrower PTW rule misses some manually confirmed cases. \ggshield has perfect precision for SEW, but a recall of 0.25, detecting only one of the four cases of secret exposure. This is consistent with its  strategy focused on exposed secret material rather than broader secret-handling weaknesses.}

\begin{tcolorbox}[boxrule=1pt,arc=.3em, left=4pt, right=4pt]
  \textbf{Answer to RQ3}: 
\rev{Based on a ground-truth of  448 weaknesses in \nbsample \ghaworkflows, the accuracy analysis confirms wide design differences among the scanners. \poutine and \frizbee are the two scanners with the lowest ratio of false positives over the weakness classes that they cover. On the other hand, the general-purpose \zizmor detects more true positives, at the cost of more false positives. These trade-offs must be carefully considered by developers who choose a security scanner.}
\end{tcolorbox}

\subsection{\researchquestionthree}

\rev{We measure the runtime performance of each scanner by executing it on each workflow and recording the median elapsed wall-clock time across three runs.}

\begin{figure}[!htbp]
\centering
\includegraphics[width=0.7\columnwidth]{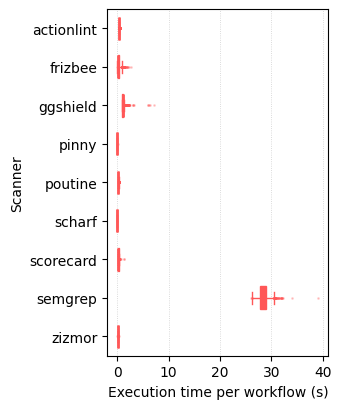}
\caption{\rev{Distribution of per-workflow execution times for each scanner.}}
\label{fig:execution_time_plot}
\end{figure}

\rev{\autoref{fig:execution_time_plot} shows the distribution of execution times for each security scanner across the \ghaworkflows.}
For each (scanner, workflow) pair, we repeat the measurement three times and report the median elapsed wall-clock time.
\rev{Each box therefore summarizes the distribution of these per-workflow median execution times for one scanner.}
This end-to-end time includes scanner startup and initialization costs, as well as the time spent analyzing the workflow.

\subsubsection{General-purpose workflow security scanners}

\actionlint, \poutine and \zizmor have stable runtimes across the dataset. This is consistent with the design of these scanners, whose rules focus on targeted workflow checks and rely on efficient static analysis implemented in Go or Rust.
% This is consistent with these scanners' design, which rules are clearly focused on targeted workflow checks and rely on efficient static analysis in Go or Rust. 
\rev{\scorecard is also fast overall, with a runtime distribution comparable to the other tools in this group.} This is a consequence of its broader repository-level checks in addition to workflow-specific analysis. \rev{Even so, all four workflow security scanners remain practical, with median per-workflow execution times of 0.40s for \actionlint, 0.22s for \poutine, 0.11s for \zizmor, and 0.25s for \scorecard.} 

\rev{Our four general-purpose workflow security scanners perform well across the dataset.} \actionlint, \poutine and \zizmor perform very similarly and are all ready for continuous integration. \scorecard introduces a limited overhead.

\subsubsection{Focused workflow security scanners}

\frizbee, \pinny and \scharf all focus on unpinned dependency weaknesses, but they do not exhibit the same runtime profiles. \rev{\pinny and \scharf are the fastest scanners in our study, with median per-workflow runtimes of 0.00s for both scanners at the precision of our measurements.} \rev{\frizbee is systematically slightly slower, with a median runtime of 0.205s, and shows a few slower outliers.} This difference is consistent with the additional timing measurements we collected: compared with \pinny and \scharf, \frizbee appears to spend more time in startup and initialization on each run, and to perform extra work to resolve external references such as commit SHAs and container image digests, rather than simply parsing the workflow file itself. As a result, the three scanners remain in the same tier overall.

\ggshield occupies an intermediate position in our setup. Its runtime is higher than that of the focused pinning scanners and the fastest general-purpose tools, but it remains well below \semgrep. This behavior is consistent with its detection strategy. \ggshield performs secret detection using entropy analysis, pattern matching, and heuristic rules. It analyzes workflow contents to identify \texttt{high-entropy} strings assigned to sensitive variable names (e.g., \texttt{token}, \texttt{secret}, or \texttt{api\_key}), applying multiple validation steps to reduce false positives. This layered content inspection contributes to variations in runtime across workflows, depending on the presence and structure of secret-like strings. This broader content inspection gives \ggshield a higher baseline cost, while still keeping runtime moderate overall. \rev{In our measurements, its median per-workflow runtime is 1.03s, with a small number of slower cases.}

\semgrep stands apart as the slowest scanner in our setup. \rev{Its runtime distribution is clearly separated from all other tools.} \rev{Its median per-workflow elapsed time is 28.31s, far above the other scanners.} This profile is consistent with a scanner that performs broader inspection of workflow contents, including shell commands and interpolated expressions, which can require deeper analysis than narrowly scoped checks. In particular, dynamic constructs such as multi-line shell commands, e.g. \texttt{docker stop \$running\_containers}, and secret references, e.g. \texttt{\${{vars.ENTERPRISE\_SSH\_\allowbreak SCRIPT || secrets.\allowbreak ENTERPRISE\_SSH\_\allowbreak SCRIPT }}}, are plausible sources of additional analysis effort. In practical terms, \semgrep is the most expensive scanner in our per-workflow execution setting.

% \todo{one short paragraph to explain why semgrep, ggshield and frizbee have these variations}

\begin{tcolorbox}[boxrule=1pt,arc=.3em, left=4pt, right=4pt]
  \rev{\textbf{Answer to RQ4}: The workflow security scanners separate into three practical runtime profiles. \semgrep is consistently the slowest scanner per workflow, \ggshield occupies an intermediate position, and the remaining scanners are fast overall, with median runtime below 0.5s.} Most workflow security scanners can be embedded as part of a continuous integration pipeline.
\end{tcolorbox}

% \begin{table}
% \footnotesize
% \caption{Comparison of \frizbee, \pinny, and \scharf with respect to their handling of unpinned GitHub Actions.}
% \label{tab:frizbeevs}
% \centering
% \rowcolors{2}{gray!10}{white}
% \begin{tabular}{lccc}
% \toprule
%  & \frizbee & \pinny & \scharf \\
% \midrule
% \makecell[l]{Skip unpinned actions referencing main or \\ master branches}
%     & \checkmark &  & \\
% Skip already pinned actions
%     & \checkmark & \checkmark & \\
% Skip commented unpinned actions
%     & \checkmark &  & \checkmark \\

% \bottomrule
% \end{tabular}
% \end{table}

\subsection{Recommendations for GitHub Actions Workflow Scanning}

% \todo{one paragraph to say that \poutine is the best among the general purpose scanners}

% \todo{one paragraph to discuss the single-purpose tools}
Our results show that there are two categories of scanners: general-purpose scanners that cover a broad set of weaknesses, and single-purpose scanners that specialize in detecting a specific type of weakness and often provide targeted fixes. As most evaluated scanners introduce limited runtime overhead, combining complementary scanners can remain practical for continuous integration pipelines. \rev{Based on  scanner coverage, accuracy, remediation support, and execution cost, we recommend three levels of scanner configurations. They can be set up in GitHub Actions workflows to achieve a fundamental level of security (\michelin), or more comprehensive levels of security scanning (\michelin \michelin or \michelin \michelin \michelin).}

\michelin \textbf{Fundamental  Scanning}

As a fundamental of \ghaworkflows scanning, we suggest adding \actionlint and \poutine to all workflows. \rev{This decision is based on the complementary capabilities of the two scanners. \poutine covers several security-oriented weakness classes using a relatively restrictive reporting strategy, while \actionlint contributes structural and GitHub runner compatibility checks that are not provided by \poutine.} \rev{The combination of these two tools provides a sound baseline for \ghaworkflows scanning. Additional scanners with higher recall can improve coverage for specific weakness classes.}

In practice, developers who wish to introduce these scanners in their repositories can set up a dedicated CI job that runs on pull requests (and optionally on pushes to the main branch) and scans the files under \texttt{.github/workflows/}. The job executes \actionlint first to catch workflow authoring and specification issues, then runs \poutine to flag security-oriented workflow patterns, as shown in \autoref{lst:recommedation}. We recommend to configure this scanning job to fail on findings so that workflow changes are merged only after the checks pass. 

\begin{center}
\begin{minipage}{\linewidth}
\begin{minted}[bgcolor=graybg, fontsize=\footnotesize, frame=single, breaklines, baselinestretch=1.2]{yaml}
name: Workflow checks (actionlint + poutine)
on: [pull_request]

jobs:
  scan-workflows:
    runs-on: ubuntu-latest
    steps:
      - uses: actions/checkout@v4
      - name: actionlint
        run: actionlint .github/workflows/*.yml
      - name: poutine
        run: poutine scan .github/workflows
\end{minted}
\captionof{listing}{Example workflow running actionlint and poutine on pull requests.}
\label{lst:recommedation}
\end{minipage}
\end{center}

\michelin \michelin \textbf{Essential Scanning}

\rev{For a more advanced level of scanning, we recommend adding a dedicated UDW scanner. \frizbee, \pinny, and \scharf all specialize in detecting unpinned dependencies, but the accuracy analysis reveals different trade-offs among them. \frizbee achieves perfect precision with lower recall, while \pinny and \scharf achieve nearly the same precision with higher recall. All three scanners also provide automated fixes for pinning dependencies, while \frizbee additionally supports both individual workflow files and full repositories and provides configurable exclusions. The choice among these scanners therefore depends on whether developers prioritize stricter reporting and remediation support or broader detection of unpinned dependencies.}

\michelin \michelin \michelin \textbf{Advanced  Scanning}

For the most specialized and comprehensive level of workflow scanning, we recommend adding \semgrep. \rev{Compared to \poutine, \semgrep contributes additional rules for detecting potential code-injection patterns.} However, it comes with substantially higher runtime overhead than the other evaluated scanners, and we therefore recommend it for scenarios where its additional checks justify this cost. \rev{This combination provides broader complementary coverage, but should not be interpreted as a complete solution, since our accuracy analysis shows that scanners can still miss weaknesses within the classes they support.}

\section{Discussion}

\subsection{Scope of the analysis}
\label{sec:discussion-scope}

\rev{It is important to interpret our results within the scope of our benchmark. We compare security scanners that are locally executable, using fixed tool versions, documented commands, and normalized outputs over the same workflow dataset. Runtime monitors, cloud-based scanners, and broader tools such as Trivy, CodeQL, or ARGUS-style analyses are complementary, but they target different artifacts, execution models, or assumptions and are therefore not directly comparable to our setup.}

\rev{The dataset scope follows the same rationale. We intentionally focus on repositories from large technology companies to obtain mature, high-impact, and structurally diverse GitHub Actions workflows. This choice gives us a meaningful benchmark for comparing scanner behavior across real-world CI/CD configurations, rather than a random sample of the entire GitHub ecosystem. At the same time, these organizations may follow stricter engineering and security practices than smaller or less security-conscious projects. Consequently, the observed weakness distributions and scanner reporting behavior should be interpreted within this curated benchmark, not as a universal distribution of all GitHub Actions workflows.}

\subsection{Criticality Levels of Reported Weaknesses}

% \todo{prepare a new version of this subsection to focus on criticality and emphasize that there currentely exists no scoring mechanism for GHA weaknesse and this should be done in futurue work}

\rev{The detection volume reported in RQ2 should not be interpreted as a direct measure of criticality or security impact. Static-analysis warnings are not equally critical: some correspond to security-relevant weaknesses, while others are closer to hardening gaps or contextual risk indicators. This distinction is especially important for GitHub Actions workflows, where there is currently no generally adopted scoring model for ranking workflow weaknesses across scanners by severity or practical security relevance.}

\rev{Some weakness classes in our taxonomy are therefore more critical than others. For example, injection weaknesses and privileged-trigger weaknesses may expose clearer attack paths when attacker-controlled workflow inputs, pull request metadata, or artifacts influence commands executed in a privileged context. 
Other reported weaknesses should be interpreted as lower or context-dependent criticality signals. For example, artifact-integrity issues, runner-compatibility problems, or missing hardening mechanisms do not always provide an immediate attack path by themselves, but they can increase the impact or likelihood of exploitation when combined with another weakness.
Future work should therefore define a scoring method for GitHub Actions workflow weaknesses that helps compare findings across scanners according to their severity and practical security relevance.}

% \todo{discuss the fact that some weaknesses are more critical than others; but more importantly, we need more research on demonstrating the 'exploitability' of weaknesses}
% \todo{triage the warnings-like reachability}

\subsection{Threats to Validity}

% The first threat to validity relates to the size of our set of case studies. 
% Our study is based on \nbworkflows workflows, and some security scanners might behave differently with other workflows. We mitigate this threat, by selecting real-world workflows with different sizes and scopes from reputable GitHub organizations. 
% The second threat relates to the absence of ground truth. 
% To the best of our knowledge, there is no verified dataset of GitHub Actions workflows labeled with known weaknesses, which can be used as a ground truth to assess the correctness of the weaknesses reported by different security scanners. We mitigate this threat with a comparative study and manual inspection of the scanners' code to explain the differences in the number of reported weaknesses.  
% The third threat relates to the correctness of our code and data analysis pipeline. We have developed an automatic pipeline to run all the scanners against \nbworkflows workflows, collect logs and reports, analyze and consolidate the number of weaknesses. This pipeline is subject to potential bugs. We mitigate this threat through testing and manual inspections. The pipeline and  experimental data is publicly available, see \autoref{sec:data}.

\rev{
\textit{Internal validity:} The first threat relates to internal validity, specifically the correctness of our code and data analysis pipeline. We developed an automatic pipeline to run all the scanners against \nbworkflows workflows, collect logs and reports, and analyze and consolidate the number of weaknesses. This pipeline is subject to potential bugs that could affect our results. We mitigate this threat through testing and manual inspections. The pipeline and experimental data are publicly available; see \autoref{sec:data}.
}

\rev{
\textit{Construct validity:} The second threat relates to construct validity, particularly the absence of an established ground-truth dataset. To the best of our knowledge, there is no verified dataset of GitHub Actions workflows labeled with known weaknesses that can be used as a ground-truth to assess the correctness of the weaknesses reported by different security scanners. We mitigate this threat with a manually curated set of \nbsample (\prctsample) workflows, for which we manually label the security weaknesses and compare them with the scanner outputs. The manual labeling may also involve subjective decisions for context-dependent weaknesses; we mitigate this risk through independent review by two team members and adjudication by a third team member in case of disagreement. We also comparatively study and inspect the source code of the scanners to help explain the differences in the number of reported weaknesses. 
However, the manually labeled sample provides uneven coverage across weakness classes: some weaknesses occur only a few times, and CFW does not occur in the sample. Therefore, the reported accuracy results should be interpreted as an assessment on this validation sample rather than as definitive estimates of scanner accuracy across all GitHub Actions workflows.
}

\rev{
\textit{External validity:} The third threat relates to the generalization of our findings beyond the benchmark. As discussed in \autoref{sec:discussion-scope}, our dataset intentionally focuses on workflows from large technology companies. This gives us mature and structurally diverse CI/CD configurations, but workflows from smaller projects or less security-conscious organizations may exhibit different weakness distributions and scanner reporting patterns. Therefore, our findings should be interpreted as a comparative analysis of scanner behavior on this curated benchmark, rather than as a universal characterization of all GitHub Actions workflows.
}

\rev{\textit{Reliability and reproducibility:} The reported findings depend on the specific scanner versions and configurations used in our benchmark. Scanner behavior may change as rules evolve. We mitigate this threat by pinning and documenting the exact commit of each scanner, executing the tools locally under a controlled setup, and documenting the exact command used for each scanner. Runtime measurements also depend on the experimental machine, whose hardware configuration we report to support interpretation and reproducibility. To reduce run-to-run variability, we repeat each measurement three times and report the median. The pipeline, scanner configurations, and collected data are publicly available; see \autoref{sec:data}.}

\section{Related work}

%CI/CD Adoption and Developer Challenges
Several studies provide context on how \ghas are used and perceived by developers. Ghaleb \etal \cite{10.1145/3736758} studied thousands of mobile projects’ pipelines and found considerable diversity and complexity in CI configurations, with frequent updates needed to keep workflows working smoothly. 
%Although this work is not directly centered on security, Section 2 provides a broad overview of CI/CD practices that helps contextualize our study, while our contribution explicitly extends this type of empirical investigation into the security domain.
Rostami \etal \cite{rostami2022use} reported that about 44\% of a large sample of GitHub projects use Actions workflows, often relying on a small core of reused Actions, and noted potential issues in versioning and consistency. 
However, many projects still underutilize CI/CD security features. 
\rev{Kubo \etal \cite{kubo2026action} further show that the adoption of \ghas security practices remains limited in practice, with implementation rates varying widely across practices and developers reporting barriers such as lack of awareness, misconceptions about applicability, and operational costs.}
Ayala \etal \cite{ayala2023workflow} found that only a minority of top projects enable automated security checks or defined disclosure policies. 
The complexity of writing workflows has led to tools like GH-WCOM  \cite{mastropaolo2024completion}, which aim to assist developers by suggesting workflow steps, and researchers have analyzed community discussions to catalog common pain points \cite{zhang2024discussion}. 
Marof and Sayed \cite{marof2024exploring} performed a thematic analysis of nearly a thousand posts and identified security vulnerabilities and dependency issues as the most prevalent concerns among \ghas users. In contrast to these works focusing on adoption and developer experience, our contribution relates to the security aspect. We specifically study the spectrum of  security weaknesses in \ghaworkflows and assess to what extent state-of-the-art workflow scanners  search for and detect those weaknesses.

%Workflow Maintenance and Quality Issues

% On the outdatedness of workflows in the \ghas ecosystem \cite{decan2023outdatedness}

% The Hidden Costs of Automation: An Empirical Study on \ghas Workflow Maintenance, SCAM 2024 \cite{valenzuela2024hidden}

% Why Do \ghas Workflows Fail? An Empirical Study \cite{zheng2025whyfail}

% Catching smells in the act: A \ghas workflow investigation, SCAM 2024 \cite{khatami2024smells}

Other related work investigates the reliability and maintainability of GitHub Actions workflows.
Decan \etal \cite{decan2023outdatedness} found that workflows frequently become outdated, referencing obsolete actions or APIs that break builds and require developer intervention. 
The maintenance burden is high \cite{valenzuela2024hidden}, with constant churn and many commits fixing CI errors, updating environments, or adapting to upstream changes. Rostami Mazrae \etal \cite{rostami5369484empirical} similarly report that workflow files are updated regularly, with many changes targeting job and step definitions or their configuration, reflecting sustained maintenance activity in practice. Zheng and colleagues \cite{zheng2025whyfail} categorized common failure causes (network timeouts, configuration mistakes, dependency breakage), showing that robust CI is non-trivial. Quality ``smells'' have been cataloged by Khatami \cite{khatami2024smells}, including seven recurring anti-patterns (e.g., broad triggers, mismanaged caching, unpinned versions) that degrade efficiency or maintainability and can indirectly affect security. These works target correctness and upkeep whereas our study centers on security, evaluating how well scanners catch weaknesses like unpinned versions in practice.

%Security Risks in GitHub Actions and CI/CD Pipelines

% \cite{pan24} identifies unpinned dependencies and injection issues as two of the most critical risks in \ghaworkflows. Our study confirms that these two issues are among the most commonly supported issues by workflow security scanners.

% Characterizing the Security of Github {CI} Workflows \cite{igibek2022characterizing}

% Continuous intrusion: Characterizing the security of continuous integration services, IEEE S\&P 2023 \todo{this paper is about CI services, maybe it's not related to our work} \cite{gu2023intrusion}

% More Haste, Less Speed: Cache Related Security Threats in Continuous Integration Services, IEEE S\&P 2024 \cite{gu2024haste}

% Toward Understanding the Security of Plugins in Continuous Integration Services CCS 2024 \cite{li2024understanding}

% Mitigating Security Issues in \ghas, EnCyCriS/SVM 2024 \cite{delicheh2024securityissues}

% Advancing DevSecOps in SMEs: Challenges and Best Practices for Secure CI/CD Pipelines, \cite{Cheenepalli2025sme} ARES 2023?

% A Comparative Study of Software Secrets Reporting by Secret Detection Tools, ICSE 2024 \cite{basak2023comparative}

 Muralee and colleagues \cite{muralee2023argus} introduced static taint analysis for GitHub Actions workflows to trace untrusted inputs and catch command-injection paths pre-execution. Rule-based scanners \cite{bendetti2022assessment} encode vulnerability and misconfiguration checks, finding  thousands of issues even in small samples. Beyond injection-oriented analyses, COSSETER \cite{tystahlcosseter} proposes demand-driven static analysis to infer the minimal permissions needed by workflows and actions, supporting least-privilege configuration rather than only flagging misconfigurations. Granite \cite{moazen2025granite} further advances least-privilege for GitHub Actions by offering a mechanism to enforce more granular permissions use during workflow execution. Reusable actions often ship weaknesses (outdated dependencies with CVEs or overly broad permissions) that propagate to consumers \cite{koishybayev2024quantifying}. Progress measurements show partial but incomplete adoption of recommended practices \cite{huang2025revisiting}. Our study contributes to the \ghaworkflows security body of knowledge with a systematic analysis of workflow security scanners, providing software development teams with concrete evidence about the strength of each scanner.

With the rise of supply-chain attacks, recent work focuses on CI workflow security. 
Pan and colleagues \cite{pan24} highlight unpinned dependencies and command injection as critical vulnerabilities in GitHub Actions. Koishybayev \etal \cite{igibek2022characterizing} analyzed 447K workflows, finding 99.8\% ran with excessive privileges, could be triggered by untrusted pull requests, and most of them used third-party actions with potential issues. Other CI platforms show isolation and privilege weaknesses \cite{gu2023intrusion}; cache abuse enables data exfiltration and artifact poisoning \cite{gu2024haste}; and vulnerable or malicious plugins threaten pipeline integrity \cite{li2024understanding}. Guidance recommends pinning versions, minimal permissions, author verification, and secret scanning \cite{delicheh2024securityissues}. Basak et al. \cite{basak2023comparative} compare secret detection scanners and report meaningful differences in what they detect and how consistently they flag leaks, which affects the practical protection secret scanning provides in CI/CD.
We contribute to this area with the first systematic differential analysis of state-of-the-art security scanners for \ghaworkflows.

\section{Conclusion}

\rev{We conduct the first systematic analysis of locally executable static security scanners for \ghas workflows.} We  curate  \nbscanner state-of-the-art scanners that statically analyze \ghaworkflows. We classify the rules implemented by these tools into 10 common types of security weaknesses in \ghaworkflows, which serve as a baseline for comparing the scanners. \rev{We run all scanners against \nbworkflows real-world workflows and evaluate their coverage, detection consistency, accuracy, and performance.}
\rev{Our results provide evidence that no single scanner performs uniformly well across all common weakness types. Scanners targeting the same weaknesses can still exhibit substantially different precision and recall.} Instead, some scanners are specialized for particular categories of weaknesses, whereas others function as more general-purpose linters for \ghaworkflows.
We also find that most scanners  run fast enough to be used in CI/CD pipelines.

Based on our findings, we recommend three levels of scanner configuration, corresponding to fundamental, essential, and advanced levels of security. For fundamental scanning, we suggest configuring \poutine and \actionlint, which provide fast, security-oriented feedback about the workflows. \rev{For essential scanning, we recommend adding a dedicated unpinned-dependency scanner, with \frizbee, \pinny, and \scharf offering different trade-offs between precision, recall, and remediation support. For advanced scanning, we advise complementing this configuration with \semgrep for additional code-injection checks.}
GitHub itself is currently strengthening the built-in security checks for \ghas Workflows, for instance, by making releases and tags immutable and by allowing organizations to enforce that GitHub Actions are pinned.
\rev{For future work, we aim to extend our manually labeled sample into a larger and more balanced ground-truth benchmark of \ghaworkflows. This would support more robust evaluation of scanner precision and recall, as well as future research on weakness detection and repair.} We also foresee fantastic opportunities in improving the scanners with the analysis of the deeper network of transitive GitHub Action reuse, as current scanners primarily focus on weaknesses in the first level of reused actions.

\section{Data Availability}
\label{sec:data}

Data and code for this work is available at \url{https://github.com/sparkrew/github-actions-security}

\section*{Acknowledgements}
This research was undertaken thanks to funding from the Canada Research Chairs Program, FRQ and NSERC.

%\balance
\bibliographystyle{ieeetr}
\bibliography{biblio}

\end{document}